\documentclass[12pt]{article}
\usepackage[margin=1in]{geometry}  
\usepackage{graphicx}
\usepackage{amsmath}
\usepackage{algorithm}
\usepackage{pdflscape}
\usepackage{cite}
\usepackage{color}
\usepackage{natbib}
\usepackage{bm}
\usepackage{amssymb}
\usepackage{epstopdf}

\def \E{\mbox{E}}
\def \Tr{\mbox{Tr}}

\def \Cov{\mbox{Cov}}
\newcommand{\bu}{{\bm u}}
\newcommand{\bg}{{\bm g}}
\newcommand{\bv}{{\bm v}}

\newcommand{\ba}{{\bm a}}

\newcommand{\bx}{{\bm x}}
\newcommand{\bX}{{\bm X}}
\newcommand{\by}{{\bm y}}
\newcommand{\bW}{{\bm W}}
\newcommand{\bV}{{\bm V}}

\newcommand{\bU}{{\bm U}}
\newcommand{\bB}{{\bm B}}
\newcommand{\bP}{{\bm P}}
\newcommand{\bz}{{\bm z}}

\newcommand{\bbeta}{{\bm \beta}}

\newcommand{\bmu}{{\bm \mu}}
\newcommand{\btheta}{{\bm \theta}}
\newcommand{\bepsilon}{{\bm \epsilon}}
\newcommand{\bone}{{\bm 1}}
\newcommand{\bzero}{{\bm 0}}
\newcommand{\bI}{{\bm I}}
\newcommand{\bA}{{\bm A}}
\newcommand{\bD}{{\bm D}}

\makeatletter
\renewcommand\section{\@startsection {section}{1}{\z@}
                                   {-3.5ex \@plus -1ex \@minus -.2ex}
                                   {2.3ex \@plus.2ex}
                                   {\normalfont\large\bfseries}}
\renewcommand\subsection{\@startsection{subsection}{2}{\z@}
                                     {-3.25ex\@plus -1ex \@minus -.2ex}
                                     {1.5ex \@plus .2ex}
                                     {\normalfont\normalsize\bfseries}}
\makeatother

\newtheorem{theorem}{Theorem}[section]
\newtheorem{lemma}[theorem]{Lemma}
\newtheorem{proposition}[theorem]{Proposition}
\newtheorem{corollary}[theorem]{Corollary}
\newtheorem{fact}[theorem]{Fact}

\title{Fused Lasso Additive Model}

\begin{document}
\author{Ashley Petersen\footnote{ajpete@uw.edu}, Daniela Witten\footnote{dwitten@uw.edu}, and Noah Simon\footnote{nrsimon@uw.edu} \\ Department of Biostatistics, University of Washington, Seattle WA 98195}
\maketitle

%%%%%%%%%%%%%%%%%%%%%%%%%%%%%%%%%%%%%%%
%%%%%%%%%%%%%%%%%%%%%%%%%%%%%%%%%%%%%%%
% abstract
%%%%%%%%%%%%%%%%%%%%%%%%%%%%%%%%%%%%%%%
%%%%%%%%%%%%%%%%%%%%%%%%%%%%%%%%%%%%%%%

We consider the problem of predicting an outcome variable using $p$ covariates that are measured on $n$ independent observations, in the setting in which flexible and interpretable fits are desirable. 
We propose the \emph{fused lasso additive model} (FLAM), in which each additive function is estimated to be piecewise constant with a small number of adaptively-chosen knots. FLAM is the solution to a convex optimization problem, for which a simple algorithm with guaranteed convergence to the global optimum is provided. FLAM is shown to be consistent in high dimensions, and an unbiased estimator of its degrees of freedom is proposed. We evaluate the performance of FLAM in a simulation study and on two data sets.  

%%%%%%%%%%%%%%%%%%%%%%%%%%%%%%%%%%%%%%%
%%%%%%%%%%%%%%%%%%%%%%%%%%%%%%%%%%%%%%%
% intro
%%%%%%%%%%%%%%%%%%%%%%%%%%%%%%%%%%%%%%%
%%%%%%%%%%%%%%%%%%%%%%%%%%%%%%%%%%%%%%%

\section{Introduction}
In this paper, we consider the task of predicting a response variable using $p$ features measured on $n$ independent observations. 
 Approaches for this task typically offer either interpretability but limited flexibility (for instance, linear regression or a piecewise constant model with pre-specified knots) or flexibility but limited interpretability (for instance, a non-parametric approach). In this paper, we propose a method that balances the trade-off between interpretability and flexibility, while also allowing for sparsity in high dimensions when $p>n$. 
It selects a subset of features to include in the model, and for these features it  fits piecewise constant functions with knots that are chosen $\textit{adaptively}$ based on the data.

We now introduce some notation. We let $\bX$ denote an $n\times p$ matrix, for which $\bx_j$ is the $j$th column (feature),  and for which the $i$th  element (observation) is $x_{ij}$. When we consider the case of $p=1$, we use $\bx$ to denote the single feature, with $i$th element $x_i$.  The response is an  $n$-vector $\by$, with $i$th element $y_i$. To reference subvectors and submatrices, we use 
$\ba_\mathcal{S}$ ($\bA_\mathcal{S}$) to denote $\ba$ ($\bA$) with only the elements (columns) contained in the set $\mathcal{S}$.

The rest of this paper is organized as follows. In Section~\ref{sec:previous}, we review related work. In Sections~\ref{sec:methods} and \ref{sec:properties}, we propose our method, present an algorithm to implement it, and examine some of its properties. Sections~\ref{sec:sim} and \ref{sec:data} contain the results of a simulation study and the analyses of two data sets.   We consider some extensions in  Section~\ref{sec:ext}, and we close with a discussion  in Section~\ref{sec:discussion}.  Proofs are in the Appendix. 

%%%%%%%%%%%%%%%%%%%%%%%%%%%%%%%%%%%%%%%
%%%%%%%%%%%%%%%%%%%%%%%%%%%%%%%%%%%%%%%
% Previous work
%%%%%%%%%%%%%%%%%%%%%%%%%%%%%%%%%%%%%%%
%%%%%%%%%%%%%%%%%%%%%%%%%%%%%%%%%%%%%%%

\section{Previous Work}
\label{sec:previous}

Generalized additive models (GAM) provide a flexible and general framework for modeling a response in low dimensions ($n>p$). 
We assume $\E[y_i | \bx_i]=g\left(\sum_{j=1}^p f_j (x_{ij})\right)$, where $g$ is a specified function and each $f_j$ is an unknown function that we wish to estimate \citep{hastie1986generalized}. For now, we restrict our attention to the case when $g$ is the identity function. There are a number of ways to estimate $f_j$ --- for example, we might use a smoothing or regression spline.

Flexible additive modeling in high dimensions has been an active area of research in recent years \citep{wood2014generalized, lin2006component, avalos2007parsimonious, sardy2004amlet, huang2010variable, li2014sparse, zhang2011linear}. Recently, \citet{ravikumar2009sparse}  proposed a high-dimensional  extension of GAM called sparse additive models (SpAM), which  induces sparsity in the function estimates using a standardized group lasso penalty \citep{simon2012standardization}. SpAM solves the  problem 
\begin{equation}
\label{eq:spam}
\displaystyle \underset{\bbeta_j\in\mathbb{R}^d, 1\leq j \leq p}{\mathrm{minimize}} \quad  \frac{1}{2n} \left \| \by-\sum_{j=1}^p \Psi_j \bbeta_j \right\|_2 ^2 + \lambda \sum_{j=1}^p \sqrt{ \frac{1}{n} \bbeta_j^T \Psi_j^T \Psi_j \bbeta_j},
\end{equation}
where $\bbeta_j= \left( \beta_{j1} \cdots \beta_{jd} \right)^T$ is a $d$-vector of coefficients, and $\Psi_j = \left[ \psi_{j1} \cdots \psi_{jd} \right]$ is an $n\times d$ matrix of which the columns are the $d$ basis functions used to model $f_j$. 

\citet{meier2009high} modified \eqref{eq:spam} in order to  obtain data-adaptive fits that can capture complex relationships if needed, but that otherwise are smooth. 
Their estimator is the solution to the optimization problem
\begin{equation}
\label{eq:spsmooth}
\displaystyle \underset{\bbeta_j\in\mathbb{R}^d, 1\leq j \leq p}{\mathrm{minimize}} \quad  \frac{1}{n} \left \| \by-\sum_{j=1}^p \Psi_j \bbeta_j \right\|_2 ^2 + \lambda_1 \sum_{j=1}^p \sqrt{ \frac{1}{n} \bbeta_j^T \Psi_j^T \Psi_j \bbeta_j} + \lambda_2\sum_{j=1}^p \sqrt{\bbeta_j^T\bW_j\bbeta_j},
\end{equation}
where $\bbeta_j= \left( \beta_{j1} \cdots \beta_{jd} \right)^T$ is a $d$-vector of coefficients to be estimated, $\Psi_j = \left[ \psi_{j1} \cdots \psi_{jd} \right]$ is an $n\times d$ matrix of which the columns are the cubic B-spline basis vectors of the $j$th predictor, and $\bW_j$ is a $d\times d$ matrix containing the inner products of the second derivatives of the cubic B-spline basis functions. Other variations of this sparsity-smoothness penalty have also been proposed \citep{meier2009high,buhlmann2011statistics}.

Recently, \citet{lou2014sparse} proposed the sparse partially linear additive model, which models a subset of the included features linearly and the remaining included features non-linearly using basis functions. The linear features do not need to be chosen \textit{a priori}.

%%%%%%%%%%%%%%%%%%%%%%%%%%%%%%%%%%%%%%%
%%%%%%%%%%%%%%%%%%%%%%%%%%%%%%%%%%%%%%%
% methods
%%%%%%%%%%%%%%%%%%%%%%%%%%%%%%%%%%%%%%%
%%%%%%%%%%%%%%%%%%%%%%%%%%%%%%%%%%%%%%%

\section{The Fused Lasso Additive Model}
\label{sec:methods}
The methods described in Section~\ref{sec:previous}  rely on a pre-specified set of basis functions. This limits flexibility, since the basis functions must be chosen \emph{a priori} rather than in a data-adaptive way, as well as interpretability, since for many choices of basis functions (e.g., natural cubic splines) the resulting fits can be complex and non-monotonic without clear change points.

We now propose an approach to fit an additive model in which each function is estimated to be piecewise constant with a small number of knots. While this problem is easily solved when the knots are chosen \textit{a priori}, our proposal allows the knots to be chosen adaptively. 

\subsection{The Optimization Problem}

To begin, we assume that we have a single feature $\bx$ that is ordered, i.e., $x_1 < x_2 < \cdots < x_n$. We wish to estimate an $n$-vector $\btheta=(\theta_1,\ldots,\theta_n)$, where $\E[y_i | x_i] = f(x_i) = \theta_i$. The fused lasso seeks a  piecewise constant estimate of $\btheta$ that involves a small number of knots, by solving the problem \citep{tibshirani2005sparsity}
\begin{equation}
\label{eq:problem2}
\displaystyle \underset{\btheta\in\mathbb{R}^n} {\mathrm{minimize}} \quad  \frac{1}{2}  \left \| \by- \btheta \right\|_2 ^2 + \lambda \left\| \bD\btheta  \right\| _1,
\end{equation}
where $\lambda\geq 0$ is a tuning parameter and $\bD$ is the discrete first derivative matrix, 
\begin{center}
$\bD =
\begin{pmatrix}
1 & -1 & 0 & \cdots & 0 & 0 \\
 0 & 1 & -1 & \cdots & 0 & 0 \\
\vdots  & &&  &  &   \\
0 & 0 & 0 &\cdots & 1 & -1
\end{pmatrix} $.
\end{center} 
We let $\hat\btheta$ denote the solution to \eqref{eq:problem2}. The $\ell_1$ penalty encourages   $ \lvert \hat\theta_{i-1} - \hat\theta_{i} \rvert$ to equal zero when $\lambda$ is large. The non-zero elements of $ \lvert \hat\theta_{i-1} - \hat\theta_{i} \rvert$ correspond to knots in $\hat\btheta$. Consequently, $\hat\btheta$ provides a piecewise constant fit to the data, with adaptively-chosen knots. 
 Several algorithms for solving \eqref{eq:problem2} have been proposed \citep{hoefling2010path, liu2010efficient, johnson2013dynamic}. 

We now consider  the model $\E[y_i|\bx_i] = \sum_{j=1}^p f_j(x_{ij}) = \sum_{j=1}^p \theta_{ji}$. We assume that  each  $\btheta_j$ is piecewise constant with mean zero, and we include an intercept $\theta_0$. Let $\bP_j$ denote a permutation matrix that orders $\bx_j$ from least to greatest. We can then solve the problem 
\begin{equation}
\label{eq:problem3}
\displaystyle \underset{\theta_0\in\mathbb{R}, \btheta_j\in\mathbb{R}^n, 1\leq j \leq p}{\mathrm{minimize}} \quad  \frac{1}{2} \left \| \by-\sum_{j=1}^p \btheta_j - \theta_0 \bf{1} \right\|_2 ^2 + \lambda \sum_{j=1}^p \left\|  \bD \bP_j \btheta_j \right \|_1 \quad\text{subject to }\bone^T\btheta_j=\bzero\quad\forall j.
\end{equation}

In high dimensions, we may wish to impose sparsity on the  $\btheta_j$'s, so that a given feature is completely excluded from the model. For $\lambda$ sufficiently large, we do get a sparse solution in (\ref{eq:problem3}), but this value of $\lambda$ will tend to overshrink all of the estimates for $\btheta_j$. Therefore, we consider the modified optimization problem
\begin{equation}
\label{eq:problem4}
\displaystyle \underset{\theta_0\in\mathbb{R}, \btheta_j\in\mathbb{R}^n, 1\leq j \leq p}{\mathrm{minimize}} \quad  \frac{1}{2} \left \| \by-\sum_{j=1}^p \btheta_j - \theta_0 \bf{1} \right\|_2 ^2 + \alpha\lambda \sum_{j=1}^p \left\|  \bD \bP_j \btheta_j \right \|_1 + (1-\alpha)\lambda \sum_{j=1}^p \left\|   \btheta_j \right \|_2,
\end{equation}
where $\lambda\geq0$ and $0\leq \alpha \leq 1$. Here, $\alpha$ provides a trade-off between encouraging $\hat\btheta_j$ to be piecewise constant, and inducing sparsity on the entire vector $\hat\btheta_j$ using the group lasso  \citep{yuan2006model}. We refer to the solution to (\ref{eq:problem4}) as the \emph{fused lasso additive model} (FLAM).

\subsection{An Algorithm for FLAM}

Problem \ref{eq:problem4} is convex, and so can be solved using a general-purpose  interior point method that has a per-iteration computational complexity of $\mathcal{O}(n^3 p^3)$. Here we develop a much faster algorithm using block coordinate descent to solve (\ref{eq:problem4}) \citep{tseng2001convergence}. We cycle through the features and repeatedly perform a partial minimization in a single $\btheta_j$, holding all others fixed. The solution for the partial minimization is given in Corollary~\ref{cor:soln}, which follows from a more general result presented in Section~\ref{subsec:otherpenalties}. This corollary allows us to solve the FLAM optimization problem in $\mathcal{O}(n)$ operations per feature per iteration, by leveraging an existing fused lasso solver that requires $\mathcal{O}(n)$ operations for an $n$-dimensional problem \citep{johnson2013dynamic}.  

\begin{corollary}
\label{cor:soln}
The solution to the optimization problem
\begin{equation}
\label{eq:eq1}
\displaystyle \underset{\btheta\in\mathbb{R}^n}{\mathrm{minimize}} \quad  \frac{1}{2} \left \| \by -\btheta \right\|_2 ^2 + \alpha\lambda  \left\|  \bD  \btheta \right \|_1 + (1-\alpha)\lambda \left\|   \btheta \right \|_2 
\end{equation}
is $\left(1 - \frac{(1-\alpha)\lambda}{\left\| \hat \btheta \right\|_2}\right)_+\hat\btheta$, \quad where $(u)_+= \max(u,0)$ 
and $\hat\btheta$ is the solution to
\begin{equation}
\label{eq:eq2}
\displaystyle \underset{\btheta\in\mathbb{R}^n}{\mathrm{minimize}} \quad  \frac{1}{2} \left \| \by -\btheta \right\|_2 ^2 + \alpha\lambda  \left\|  \bD  \btheta \right \|_1.  
\end{equation}
\end{corollary}
Corollary~\ref{cor:soln} leads directly to Algorithm~\ref{alg:flam}, which  yields the global optimum to (\ref{eq:problem4}) \citep{tseng2001convergence}, and  can be made very efficient using warm starts and active sets.
\begin{algorithm}
\caption{--- Block Coordinate Descent for Fused Lasso Additive Model (Equation \ref{eq:problem4})} \vskip .2 cm
\label{alg:flam}
\begin{enumerate}
\item Initialize $\hat\theta_0=0$ and $\hat\btheta_j=\mathbf{0}$ for all $j=1,\ldots,p$. 
\item For each $j=1,\ldots, p$, perform the following:
\begin{itemize}
\item[a.] Compute the residual ${\bm r}_j = \by -\hat\theta_0{\bf 1}- \sum_{j'\neq j} \hat\btheta_{j'}$.
\item[b.] Using an algorithm for the fused lasso (e.g., \texttt{flsa} on \texttt{CRAN} \citep{flsa}), solve
\begin{center}
$\displaystyle \underset{\btheta_j\in\mathbb{R}^n}{\mathrm{minimize}}\quad \frac{1}{2} \left \| \bm{r}_j - \btheta_j \right\|_2 ^2 + \alpha\lambda \left\|  \bD \bP_j \btheta_j \right \|_1$.
\end{center}
\item[c.] Compute the intercept, $\hat\theta_0\gets\hat\theta_0 + \text{mean}(\hat\btheta_j)$, and center, $\hat\btheta_j \gets \hat\btheta_j - \text{mean}(\hat\btheta_j)$.
\item[d.] Soft-scale the estimate: $\hat\btheta_j \gets \left(1 - \frac{(1-\alpha)\lambda}{\left\| \hat \btheta_j \right\|_2}\right)_+\hat\btheta_j$ where $(u)_+= \max(u,0).$
\end{itemize}
\item Repeat Step 2 until convergence of the objective of (\ref{eq:problem4}).
\end{enumerate}
\end{algorithm}

\subsection{Connections to Other Methods}

The fused lasso can be interpreted as $\ell_1$ trend filtering with order $k=0$  \citep{kim2009ell_1,tibshirani2014}. \citet{tibshirani2014} showed that  0th order trend filtering is equivalent to 0th order locally adaptive regression splines, proposed by  \citet{mammen1997locally}. Therefore, we can interpret  FLAM with $\alpha=1$ as a multi-variable extension of locally adaptive regression splines. Indeed, we illustrate FLAM's local adaptivity, or ability to produce a fit that is highly variable in one portion of the domain and constant in another, in Section~\ref{sec:sim}.

When $\alpha=0$, FLAM is equivalent to SpAM with $\Psi_j=\bI\in\mathbb{R}^{n \times n}$ in (\ref{eq:spam}). However, this is an impractical special case in which the design matrix does not depend on the covariates. 

%%%%%%%%%%%%%%%%%%%%%%%%%%%%%%%%%%%%%%%
%%%%%%%%%%%%%%%%%%%%%%%%%%%%%%%%%%%%%%%
% properties of FLAM (df, max lambda, prediction consistency)
%%%%%%%%%%%%%%%%%%%%%%%%%%%%%%%%%%%%%%%
%%%%%%%%%%%%%%%%%%%%%%%%%%%%%%%%%%%%%%%

\section{Properties of FLAM}
\label{sec:properties}

We  define $\tilde\by=\by - \frac{1}{n}\bone^T\by\bone$ and $\bV=[\bP_1^T \bU\cdots \bP_p^T \bU]$, where $\bU\in\mathbb{R}^{n\times (n-1)}$ is the matrix obtained by centering the columns of the upper triangular matrix of 1's, and removing the $n$th column. 
The following lemma indicates that FLAM  can be reparameterized in terms of the pairwise differences among the ordered elements of $\btheta_j$  (i.e., the elements of $\bbeta_{j}=\bD\bP_j\btheta_j$).

\begin{lemma}
\label{lem:lasso}
Let $\hat\bbeta = (\hat\bbeta_1^T \; \cdots \; \hat\bbeta_p^T)^T$ be the solution to
\begin{equation}
\label{eq:lasso}
\displaystyle  \underset{\bbeta\in\mathbb{R}^{(n-1)p}}{\mathrm{minimize}}\quad \frac{1}{2} \left \|\tilde\by-\bV\bbeta  \right\|_2 ^2 + \alpha\lambda \left\|  \bbeta\right \|_1+(1-\alpha)\lambda\sum_{j=1}^p\left \| \bU\bbeta_{j}\right\|_2. 
\end{equation}
Then the solution $\hat\theta_0, \hat\btheta_1, \ldots, \hat\btheta_p$ to the optimization problem (\ref{eq:problem4})
is
$$
\hat\theta_0=\frac{1}{n}\bone^T\by\text{ and }\hat\btheta_j = \bP_j^T\bU\hat\bbeta_j\text{ for }j=1,\ldots,p.
$$
\end{lemma}
From (\ref{eq:lasso}), FLAM with $\alpha=1$ is equivalent to solving a lasso problem. The reparametrization given in Lemma~\ref{lem:lasso} will allow us to easily derive some properties of FLAM. 

\subsection{Degrees of Freedom for FLAM}
\label{subsec:dfflam}
\begin{figure}[ht!]
\begin{center}
\includegraphics[width=8cm]{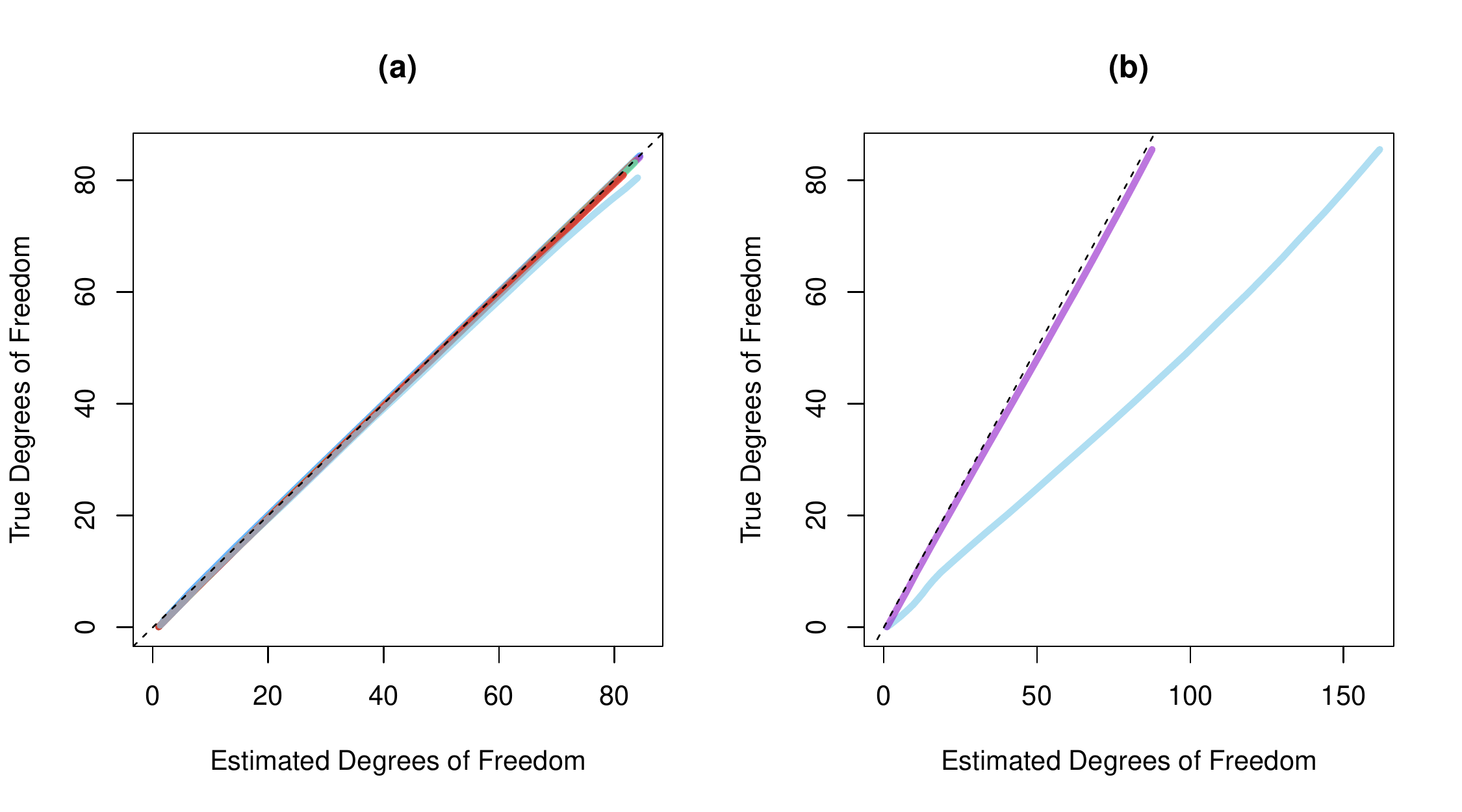}
\end{center}
\caption{In \textit{(a)}, we compare the degrees of freedom of FLAM calculated using \eqref{eq:simdf} (y-axis) to the unbiased estimator (\ref{eq:dfestflam}) (x-axis) for $\alpha=\{0,0.2,0.4,0.6,0.8,1\}$; each value of $\alpha$ is indicated by an (overlapping) colored line. In \textit{(b)}, we compare the degrees of freedom for SpAM with $d=3$  calculated using \eqref{eq:simdf} (y-axis) to the estimators \eqref{eq:dfestspam}  (\protect\includegraphics[height=.17cm]{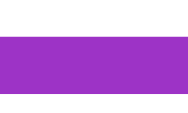}) and \eqref{ravidf} (\protect\includegraphics[height=.17cm]{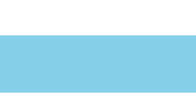}) (x-axis). In both plots, the solid lines are obtained by varying $\lambda$ for FLAM or SpAM. The black dotted lines indicate $y=x$.}
\label{fig:dof}
\end{figure}

Suppose that $\by\sim (\bmu,\sigma^2\bI)$, and let $g(\by)=\hat{\by}$ denote the fit corresponding to some model-fitting procedure $g$. Then the  degrees of freedom of  $g$ is defined as $\frac{1}{\sigma^2}\sum_{i=1}^n \Cov(y_i,\hat y_i)$ \citep{hastie1990generalized,efron1986biased}. 
We now consider a modified version of FLAM, in which a small ridge penalty  ensures strict convexity and enforces uniqueness of the solution, 
\footnotesize
\begin{equation}
\label{eq:ridgeform}
\displaystyle \underset{\theta_0\in\mathbb{R}, \btheta_j\in\mathbb{R}^n, 1\leq j \leq p}{\mathrm{minimize}} \quad  \frac{1}{2} \left \| \by-\sum_{j=1}^p \btheta_j - \theta_0 \bf{1} \right\|_2 ^2 + \alpha\lambda \sum_{j=1}^p \left\|  \bD \bP_j \btheta_j \right \|_1 + (1-\alpha)\lambda \sum_{j=1}^p \left\|   \btheta_j \right \|_2+\frac{\epsilon}{2}\sum_{j=1}^p \| \bD\bP_j\btheta_j\|_2^2.
\end{equation}
\normalsize
In \eqref{eq:ridgeform}, $\epsilon\geq0$ is a very small constant. Subject to reparameterization, (\ref{eq:ridgeform}) is equivalent to
\begin{equation}
\label{eq:lassoridge}
\displaystyle  \underset{\bbeta\in\mathbb{R}^{(n-1)p}}{\mathrm{minimize}}\quad \frac{1}{2} \left \|\tilde\by-\bV\bbeta  \right\|_2 ^2 + \alpha\lambda \left\|  \bbeta\right \|_1+(1-\alpha)\lambda\sum_{j=1}^p\left \| \bU\bbeta_{j}\right\|_2+\frac{\epsilon}{2}\sum_{j=1}^p \| \bbeta_j\|_2^2.
\end{equation}
We propose to estimate the degrees of freedom of FLAM as 
\begin{equation}
\label{eq:dfestflam}
\hat{df}_{FLAM} = \Tr\left( \bV_{\mathcal{A}}\left[ \bV_{\mathcal{A}}^T \bV_{\mathcal{A}}+ (1-\alpha)\lambda S_2 + \epsilon \bI \right]^{-1} \bV_{\mathcal{A}}^T  \right)+1,
\end{equation}
where $\mathcal{A}=\{i\in\{1,\ldots,(n-1)p\}:\hat\beta_{i}\neq 0\}$ 
and $S_2$ is block diagonal with the $j$th block equal to
$ \frac{\bU_{\mathcal{A}_j}^T\bU_{\mathcal{A}_j}}{\| \bU_{\mathcal{A}_j}\hat\bbeta_{j\mathcal{A}_j} \|_2} - \frac{\bU_{\mathcal{A}_j}^T\bU_{\mathcal{A}_j}\hat\bbeta_{j\mathcal{A}_j}\hat\bbeta_{j\mathcal{A}_j}^T\bU_{\mathcal{A}_j}^T\bU_{\mathcal{A}_j}}{\| \bU_{\mathcal{A}_j} \hat\bbeta_{j\mathcal{A}_j} \|_2^3}$
with $\mathcal{A}_j=\{i\in\{1,\ldots,n-1\}:\hat\beta_{ji}\neq 0\}$. 
\begin{proposition}
\label{prop:df}
Assume $\by\sim MVN(\bmu,\sigma^2\bI)$. Then $\hat{df}_{FLAM}$ is an unbiased estimator of the degrees of freedom of  FLAM.
\end{proposition}
 When $\alpha=1$ and $\epsilon=0$, (\ref{eq:dfestflam}) reduces to $\text{rank}(\bV_\mathcal{A}) + 1$, which agrees with the estimator  proposed in \citet{tibshirani2012degrees}. Recall that $\hat\bbeta_j = \bD\bP_j\hat\btheta_j$, so $\hat\beta_{ji}$ is the difference between $[\bP_j\hat\btheta_j]_{i-1}$ and $[\bP_j\hat\btheta_j]_i$. Thus non-zero elements of $\hat\bbeta$ correspond to knots in the estimated fits. When $\bV_\mathcal{A}$ is full rank (which only occurs when the number of knots is smaller than $n$), the following corollary provides a simple estimator for FLAM's degrees of freedom. 
\begin{corollary}
\label{cor:df}
Suppose that $\bV_\mathcal{A}$ is full rank, and let $\alpha=1$ in \eqref{eq:ridgeform}. Then the degrees of freedom of FLAM  is one greater than the total number of knots across all estimated fits.
\end{corollary}
In 1000 replicate data sets, we compare the mean of (\ref{eq:dfestflam}) to the mean of
\begin{equation}
\label{eq:simdf}
\frac{1}{\sigma^2}\sum_{i=1}^n \left(\hat y_i-\mu_i\right)\left(y_i-\mu_i\right),
\end{equation}
which is an estimator of $\frac{1}{\sigma^2}\sum_{i=1}^n \Cov(y_i,\hat y_i)$.  Data are generated according to the high-dimensional setting of scenario 1, described in Section~\ref{sec:sim}. Results are displayed in Figure~\ref{fig:dof}(a).

We also propose an estimator for the degrees of freedom of SpAM (\ref{eq:spam}). Defining $\Psi= \left[ \Psi_{1} \cdots \Psi_{d} \right]$ and $\bbeta= \left( \bbeta_{1}^T \cdots \bbeta_{d}^T \right)^T$, we estimate SpAM's degrees of freedom as
\begin{equation}
\label{eq:dfestspam}
\hat{df}_{SpAM} = \Tr\left(\Psi_{\mathcal{A}}\left[ \Psi_{\mathcal{A}}^T \Psi_{\mathcal{A}}+ \lambda D_2\right]^{-1} \Psi_{\mathcal{A}}^T \right),
\end{equation}
where $\mathcal{A}=\{ i\in\{1,\ldots,pd\}: \hat\beta_i \neq 0\}$ and $D_2$ is  block diagonal  with the $j$th block equal to 
$ \frac{\Psi_{j\mathcal{A}_j}^T\Psi_{j\mathcal{A}_j}}{\| \Psi_{j\mathcal{A}_j} \hat\bbeta_{j\mathcal{A}_j} \|_2} - \frac{\Psi_{j\mathcal{A}_j}^T\Psi_{j\mathcal{A}_j}\hat\bbeta_{j\mathcal{A}_j}\hat\bbeta_{j\mathcal{A}_j}^T\Psi_{j\mathcal{A}_j}^T\Psi_{j\mathcal{A}_j}}{\| \Psi_{j\mathcal{A}_j} \hat\bbeta_{j\mathcal{A}_j} \|_2^3}$
with $\mathcal{A}_j=\{i\in\{1,\ldots,d\}:\hat\beta_{ji}\neq 0\}$. 
\begin{proposition}
\label{prop:dfspam}
Assume $\by\sim MVN(\bmu,\sigma^2\bI)$. Then $\hat{df}_{SpAM}$ is an unbiased estimator of the degrees of freedom of SpAM. 
\end{proposition}
Interestingly, \citet{ravikumar2009sparse} proposed  
\begin{equation}
\sum_{j=1}^p \sum_{k=1}^d I\left(\hat\beta_{jk} \neq 0\right)
\label{ravidf}
\end{equation} as an estimator for SpAM's degrees of freedom.  Figure~\ref{fig:dof}(b) compares the means of the two estimators (\ref{eq:dfestspam}) and \eqref{ravidf} 
with the mean of (\ref{eq:simdf}) across 1000 replicate data sets. In fact, we see that \eqref{eq:dfestspam} is far more accurate than \eqref{ravidf}.

\subsection{Range of $\lambda$ that Yields Complete Sparsity}
\label{subsec:tune}

We now consider the range of  $\lambda$ for which $\hat\btheta_j=\bzero$ for $j=1,\ldots,p$, for $\alpha=1$ and $\alpha=0$.
\begin{lemma}
\label{lem:sparse}
If $\alpha=1$, then the solution to \eqref{eq:problem4} is completely sparse if and only if $\lambda  \geq \|  \bV^T\tilde\by \|_\infty$. If $\alpha=0$, then the solution is completely sparse if and only if $\lambda \geq \| \tilde\by\|_2$.
\end{lemma} 
In Lemma~\ref{lem:sparse}, note that $\|  \bV^T\tilde\by \|_\infty = \max\left(g(\bP_1\tilde\by),\ldots,g(\bP_p\tilde\by)\right)$, where $g(\ba)=\max\{\lvert a_1  \rvert, \lvert a_1+a_2 \rvert, \ldots, \lvert a_1+a_2+\ldots+a_{n-1}  \rvert\}$.   
We now present a sufficient condition for the FLAM solution to be completely sparse, for any $\alpha$. 
\begin{corollary}
\label{cor:sparseall}
For any $\alpha\in[0,1]$, if 
$\lambda \geq \min \left( \frac{\|  \bV^T\tilde\by \|_\infty}{\alpha}, \frac{\| \tilde\by\|_2}{1-\alpha} \right)$, then the solution to (\ref{eq:problem4}) is completely sparse.
\end{corollary}
When selecting $\lambda$ for FLAM, we need never consider a value larger than that in Corollary~\ref{cor:sparseall}. 

\subsection{Prediction Consistency}
\label{subsec:consist}

In this section, we establish prediction consistency for FLAM. For simplicity, we assume $\by$ has mean zero in this subsection.  The estimated prediction error compares the predicted outcome to the best one could do if the true coefficient values $\btheta_1^0,\ldots,\btheta_p^0$ were known. Lemma \ref{lem:consist} provides a finite sample bound for the prediction error. 
\begin{lemma}
\label{lem:consist}
Assume $\by = \sum_{j=1}^p \btheta_j^0 + \bepsilon$ with $\bepsilon\sim MVN(\bzero, \sigma^2\bI)$. If  $\lambda\geq 2\sigma\sqrt{\frac{\log ((n-1)p)}{n}}$,  then
$$\frac{1}{n}\left\| \sum_{j=1}^p \left(\hat\btheta_j - \btheta_j^0\right)\right\|^2_2 \leq 3 \lambda \sum_{j=1}^p \left[ \alpha\left\|  \bD \bP_j \btheta^0_j \right \|_1 +(1-\alpha)\|\btheta_j^0\|_2 \right]$$
holds with probability at least $1-\left(\frac{2}{(n-1)p} + \frac{1}{n}\right)$.
\end{lemma}

Now, assume that  $\theta_{ji}^0=f_j(x_{ij})$ where  $f_j$ has bounded variation, and all elements of $\bx_{j}\in[a,b]$ for some $a$ and $b$. Assume also that the number of non-sparse functions is bounded, i.e., $\sum_{j=1}^p ||f_j||_0 = K < \infty$. Together, these two assumptions imply that $\sum_{j=1}^p \|\bD \bP_j \btheta_j^0 \|_1 = O(1)$, and that $\sum_{j=1}^p \| \btheta_j^0\|_2 = O(\sqrt{n})$. Thus, FLAM is prediction consistent provided that $(1-\alpha)=o((\log ( (n-1)p))^{-1/2})$, and $\lambda=2\sigma\sqrt{\frac{\log ((n-1)p)}{n}}$. 

%%%%%%%%%%%%%%%%%%%%%%%%%%%%%%%%%%%%%%%
%%%%%%%%%%%%%%%%%%%%%%%%%%%%%%%%%%%%%%%
% simulations
%%%%%%%%%%%%%%%%%%%%%%%%%%%%%%%%%%%%%%%
%%%%%%%%%%%%%%%%%%%%%%%%%%%%%%%%%%%%%%%

\section{Simulations}
\label{sec:sim}

\begin{figure}[ht!]
\begin{center}
\includegraphics[width=16cm]{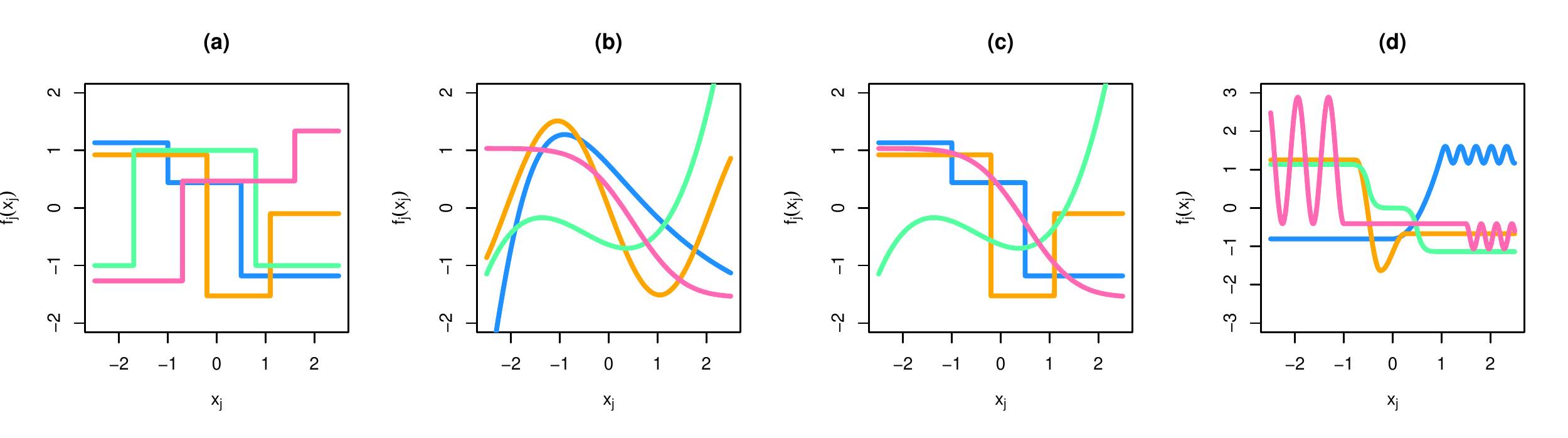}
\end{center}
\caption{Functions used to generate data in: \textit{(a)} scenario 1, \textit{(b)} scenario 2, \textit{(c)} scenario 3, and \textit{(d)} scenario 4.}
\label{fig:fj}
\end{figure}

We compare the performance of FLAM to two competitors: GAM using smoothing splines (implemented with the \texttt{R} package \texttt{gam} \citep{gam}), and SpAM with basis vectors corresponding to a natural cubic spline with $d-1$ non-boundary knots at equally spaced quantiles of $\bx_j$ (implemented with the \texttt{R} package \texttt{SAM} \citep{SAM}).  Data are generated according to $y_i=\sum_{j=1}^p f_j (x_{ij}) + \epsilon_i$ with $\epsilon_i \stackrel{iid}{\sim} N(0,1)$, $x_{ij}\stackrel{iid}{\sim} \text{Uniform}[-2.5,2.5]$, and $p=4$. We consider four scenarios, displayed in Figure~\ref{fig:fj}:

\begin{enumerate}
\item[]{\textit{Scenario 1}}: All $f_j$ are piecewise constant functions (Figure~\ref{fig:fj}(a)).

\item[]{\textit{Scenario 2}}: All $f_j$ are smooth functions (Figure~\ref{fig:fj}(b)). These are the exact functions used for the simulations in the original SpAM paper \citep{ravikumar2009sparse}.

\item[]{\textit{Scenario 3}}: Two of the $f_j$ are piecewise constant functions and the other two $f_j$ are smooth functions (Figure~\ref{fig:fj}(c)). This is a compromise between scenarios 1 and 2.

\item[]{\textit{Scenario 4}}: All $f_j$ are functions that are constant in some areas of the domain and highly variable in other areas of the domain (Figure~\ref{fig:fj}(d)).

\end{enumerate}
All functions are constructed such that $\int_{-2.5}^{2.5} f_j = 0$ and $\int_{-2.5}^{2.5} f_j^2 = 1$. We refer to scenarios 1-4 as  the \emph{low-dimensional setting}. Additionally, we refer to the same scenarios with the addition of 96 noise functions (i.e., $f_5,\ldots,f_{100}=0$)  as the \emph{high-dimensional setting}. We note that GAM can only be applied in the low-dimensional setting ($n>p$). 

For each scenario, we generate training, test, and validation sets, each with $n=100$. Functions are fit on the training set, and mean squared error 
 $\left( \textrm{MSE; } \frac{1}{n}\left \| \by -\hat \by \right\|_2 ^2\right)$ is evaluated on the test set. 
For FLAM, we fix $\alpha=0.5,0.75,\text{ or }1$ and consider a range of $\lambda$. For GAM, we consider a range of degrees of freedom for each smoothing spline, from 1 (just a linear fit) to $n/p$. For SpAM, we fix $d=3,6,\text{ or }10$ and consider a range of $\lambda$.  

We evaluate each method's performance as a function of its degrees of freedom. For GAM, the total degrees of freedom is $p$ multiplied by the degrees of freedom for each covariate's smoothing spline, plus one degree of freedom for the intercept.  For FLAM and SpAM, the degrees of freedom are estimated using (\ref{eq:dfestflam}) and (\ref{eq:dfestspam}), respectively.

\begin{figure}[ht!]
\begin{center}
\includegraphics[width=13.2cm]{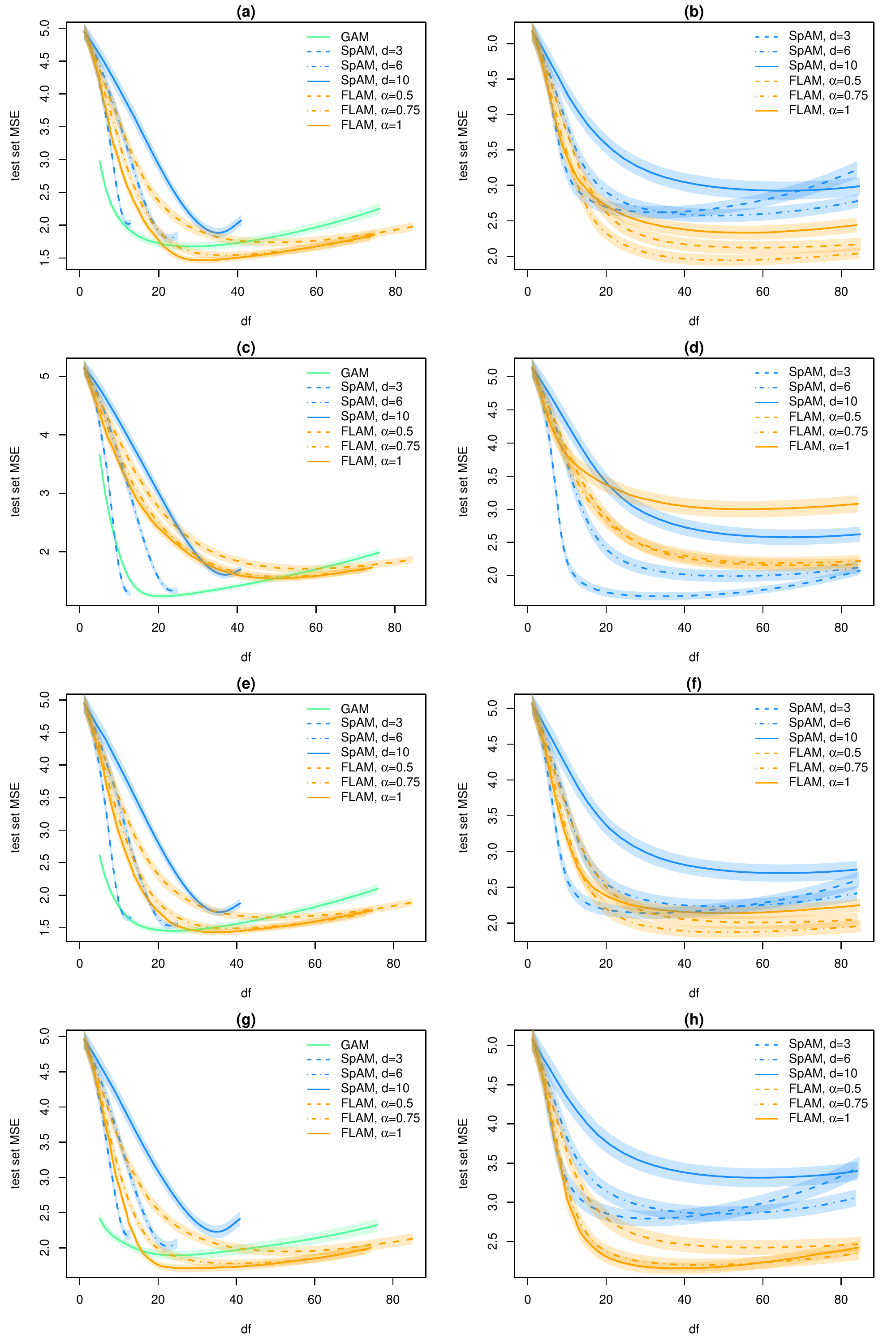}
\end{center}
\caption{Mean test set MSE plotted by degrees of freedom (df) for: \textit{(a)} scenario 1, \textit{(b)} scenario 1 with noise functions, \textit{(c)} scenario 2, \textit{(d)} scenario 2 with noise functions, \textit{(e)} scenario 3, \textit{(f)} scenario 3 with noise functions, \textit{(g)} scenario 4, and \textit{(h)} scenario 4 with noise functions. Shaded bands indicate point-wise 95\% confidence intervals over the 100 replicate data sets. GAM is only applicable in the low-dimensional setting.}
\label{fig:simbig}
\end{figure}

Figure~\ref{fig:simbig} displays the test set MSE versus total degrees of freedom for the three methods. FLAM achieves the lowest test set MSE across all scenarios except in scenario 2 where all $f_j$ are smooth. GAM performs comparably to SpAM in scenario 3 without noise functions. As expected, FLAM with $\alpha=1$ outperforms  FLAM with $\alpha<1$ in the scenarios without noise functions, as no additional sparsity is needed. In general, $\alpha<1$ is preferred in the scenarios with noise functions, with the exception of scenario 4. We discuss this discrepancy below.

Additionally, we summarize performance for the \emph{optimal} tuning parameter, defined as the tuning parameter corresponding to the minimum test set MSE. We calculate the validation set MSE for the training set fit corresponding to the optimal tuning parameter, as well as the parameter fit $( \sum_{j=1}^p  \| \btheta_j -  \hat\btheta_j \|_2 ^2)$, sparsity, and degrees of freedom (Table~\ref{tab:valdata}). Once again, FLAM performs best in all scenarios except when all $f_j$ are smooth (scenario 2), with the best performance corresponding to FLAM with $\alpha=1$ in the low-dimensional setting and $\alpha=0.75$ in the high-dimensional setting. In scenario 4 with  noise functions, FLAM with $\alpha=1$ is able to achieve comparable sparsity to FLAM with $\alpha<1$. This explains the optimal performance of FLAM with $\alpha=1$ in this setting (Figure \ref{fig:simbig}). 

\begin{table}[ht]
\caption{Results on the validation data, with the tuning parameter value chosen based on test set MSE. Mean (standard error) across 100 replicate data sets are shown. Parameter fit is defined as $ \sum_{j=1}^p \left \| \btheta_j -  \hat\btheta_j \right\|_2 ^2$. Note that GAM can only be applied in the low-dimensional setting. All $\hat f_j$ were non-zero for all methods in the low-dimensional setting.} \vskip .4 cm
\label{tab:valdata}
\centering
\tabcolsep=0.15cm
\scalebox{0.89}{
\begin{tabular}{rccc|cccc}
  \hline
  &\multicolumn{3}{c}{\textit{Low-dimensional}}&\multicolumn{4}{c}{\textit{High-dimensional}}\\
  \hline
 &  & Parameter  & Degrees of & &Parameter  & Proportion &Degrees of  \\ 
  & MSE &  fit &  freedom & MSE &  fit & $\hat f_j$ non-zero & freedom\\ 
  \hline
 \textit{Scenario 1} & & & & & & &\\
FLAM, $\alpha=0.5$ & 1.73 (0.03) & 75.6 (1.8) & 51.0 (1.1) & 2.11 (0.05) & 113.0 (3.3) & 0.20 (0.01) & 61.6 (1.5) \\ 
  FLAM, $\alpha=0.75$ & 1.52 (0.03) & 55.4 (1.5) & 39.0 (1.0) & 1.92 (0.04) & 95.0 (2.9) & 0.23 (0.01) & 54.0 (1.5) \\ 
  FLAM, $\alpha=1$ & 1.45 (0.02) & 48.2 (1.4) & 32.7 (0.8) & 2.30 (0.06) & 132.7 (3.4) & 0.35 (0.01) & 58.8 (1.9) \\ 
  GAM &1.67 (0.02) & 65.1 (1.3) & 28.7 (0.9) &  &  &  &  \\ 
  SpAM, $d=3$ & 2.02 (0.03) & 107.5 (1.4) & 12.2 (0.1) & 2.54 (0.05) & 162.0 (2.6) & 0.23 (0.01) & 36.7 (1.3) \\ 
  SpAM, $d=6$ &1.79 (0.03) & 110.0 (3.0) & 22.7 (0.2) & 2.52 (0.05) & 168.2 (3.3) & 0.23 (0.01) & 50.7 (1.5) \\ 
 SpAM, $d=10$ &1.85 (0.03) & 130.5 (3.3) & 35.4 (0.3) & 2.91 (0.06) & 208.9 (4.1) & 0.24 (0.01) & 66.5 (1.6) \\
    \hline
 \textit{Scenario 2} & & & & & & &\\
  FLAM, $\alpha=0.5$ & 1.66 (0.03) & 69.1 (1.9) & 60.0 (1.1) & 2.14 (0.05) & 112.6 (3.4) & 0.22 (0.01) & 71.4 (1.5) \\
  FLAM, $\alpha=0.75$ & 1.51 (0.02) & 56.1 (1.4) & 52.9 (0.9) & 2.17 (0.05) & 115.8 (3.4) & 0.27 (0.01) & 66.5 (1.6) \\ 
  FLAM, $\alpha=1$ & 1.46 (0.02) & 52.8 (1.3) & 50.2 (0.9) & 2.94 (0.06) & 192.5 (4.1) & 0.36 (0.01) & 60.9 (2.3) \\ 
  GAM &  1.19 (0.02) & 23.3 (0.7) & 21.6 (0.4) &  &  &  &  \\ 
  SpAM, $d=3$ & 1.21 (0.02) & 32.9 (1.3) & 12.5 (0.1) & 1.65 (0.03) & 70.3 (2.3) & 0.23 (0.01) & 37.4 (1.5) \\
  SpAM, $d=6$ & 1.27 (0.02) & 54.6 (2.1) & 23.6 (0.1) & 1.95 (0.05) & 109.7 (3.6) & 0.23 (0.01) & 53.0 (1.5) \\ 
  SpAM, $d=10$ &1.50 (0.03) & 84.1 (2.4) & 37.1 (0.2) & 2.55 (0.06) & 168.3 (4.6) & 0.25 (0.01) & 68.6 (1.5) \\ 
    \hline
 \textit{Scenario 3} & & & & & & &\\
  FLAM, $\alpha=0.5$ & 1.63 (0.03) & 66.8 (1.6) & 51.8 (1.1) & 1.98 (0.04) & 101.2 (3.3) & 0.20 (0.01) & 61.6 (1.5) \\ 
  FLAM, $\alpha=0.75$ & 1.45 (0.02) & 49.2 (1.3) & 40.2 (0.9) & 1.84 (0.04) & 88.1 (2.7) & 0.24 (0.01) & 54.9 (1.6) \\ 
  FLAM, $\alpha=1$ & 1.38 (0.02) & 43.7 (1.3) & 36.5 (0.8) & 2.12 (0.04) & 115.2 (3.1) & 0.31 (0.01) & 55.0 (2.1) \\ 
  GAM & 1.44 (0.02) & 44.2 (1.0) & 24.1 (0.7) &  &  &  &  \\ 
  SpAM, $d=3$ &1.62 (0.02) & 69.0 (1.4) & 12.2 (0.1) & 2.09 (0.04) & 115.3 (2.7) & 0.23 (0.01) & 35.1 (1.3) \\
  SpAM, $d=6$ & 1.51 (0.02) & 73.3 (2.4) & 22.9 (0.1) & 2.18 (0.04) & 134.4 (3.5) & 0.23 (0.01) & 50.6 (1.3) \\
  SpAM, $d=10$ &1.68 (0.03) & 103.7 (3.1) & 35.7 (0.2) & 2.68 (0.05) & 188.3 (4.2) & 0.24 (0.01) & 66.7 (1.6) \\ 
  \hline
 \textit{Scenario 4} & & & & & & &\\
  FLAM, $\alpha=0.5$ &1.91 (0.04) & 91.4 (1.9) & 55.6 (1.3) & 2.38 (0.05) & 137.1 (3.2) & 0.21 (0.01) & 61.5 (1.9) \\ 
  FLAM, $\alpha=0.75$ &1.73 (0.03) & 74.5 (1.6) & 42.9 (1.2) & 2.15 (0.04) & 115.4 (2.6) & 0.21 (0.01) & 45.8 (1.7) \\  
  FLAM, $\alpha=1$ & 1.64 (0.03) & 67.2 (1.5) & 33.8 (1.3) & 2.13 (0.03) & 112.1 (2.4) & 0.25 (0.01) & 43.1 (1.6) \\
  GAM & 1.88 (0.03) & 82.2 (1.4) & 28.5 (1.6) &  &  &  &  \\ 
  SpAM, $d=3$ & 2.15 (0.03) & 121.2 (2.0) & 12.2 (0.1) & 2.75 (0.05) & 176.2 (3.2) & 0.21 (0.01) & 32.9 (1.2) \\ 
  SpAM, $d=6$ &2.01 (0.03) & 120.6 (2.6) & 22.6 (0.2) & 2.78 (0.05) & 187.8 (3.9) & 0.24 (0.01) & 51.3 (1.7) \\ 
  SpAM, $d=10$ & 2.19 (0.04) & 157.1 (3.5) & 35.0 (0.3) & 3.23 (0.06) & 237.4 (4.4) & 0.22 (0.01) & 60.0 (1.8) \\
   \hline
\end{tabular}
}
\end{table}

\begin{figure}[ht!]
\begin{center}
\includegraphics[width=16cm]{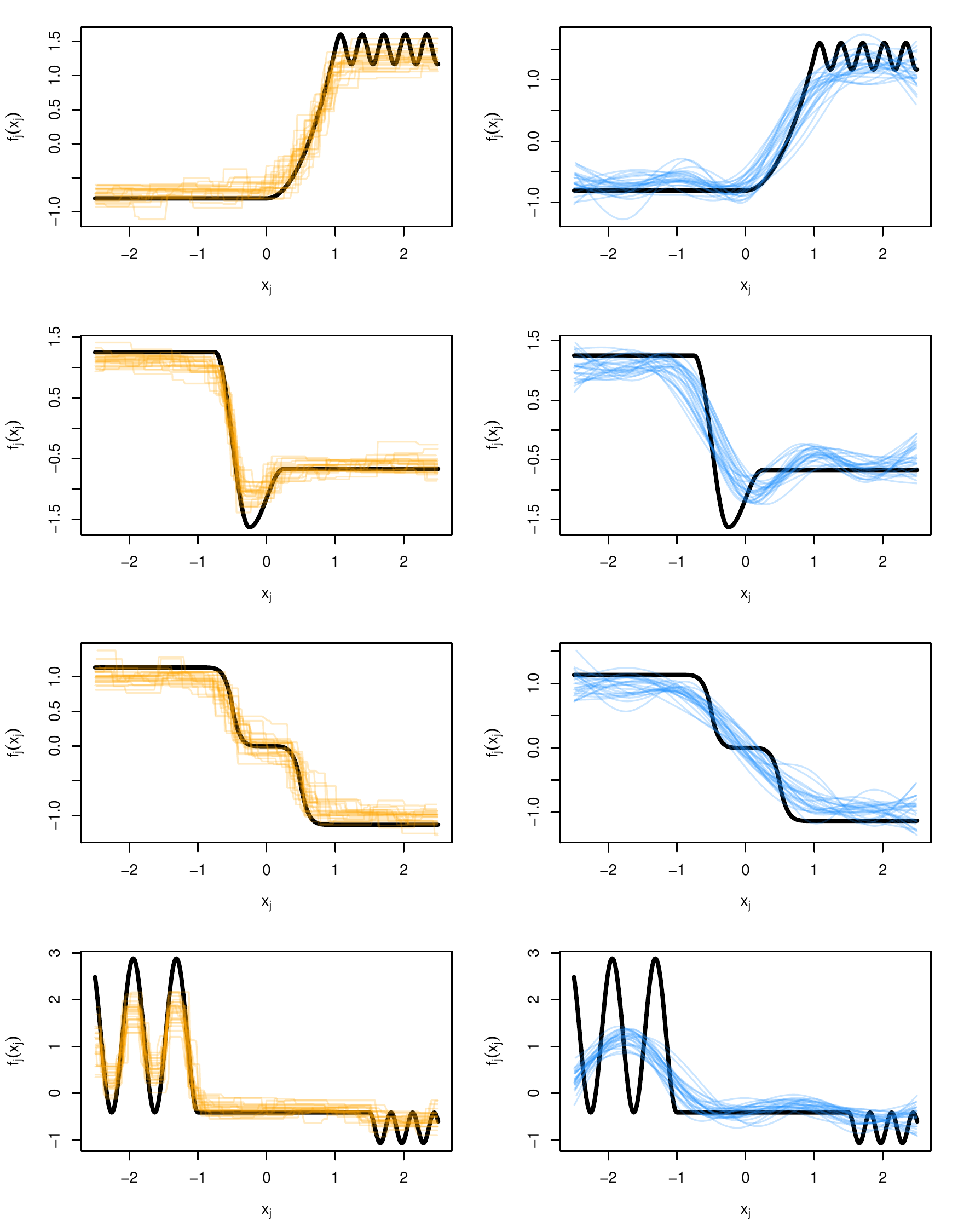}
\end{center}
\caption{We compare the fits of FLAM (\protect\includegraphics[height=.15cm]{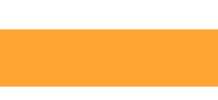}) and SpAM (\protect\includegraphics[height=.15cm]{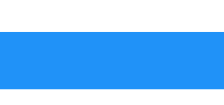}) with the true underlying functions (\protect\includegraphics[height=.15cm]{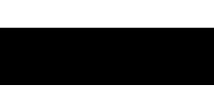}) generated according to scenario 4. In each panel, 25 curves are shown. Each curve corresponds to a validation set fit with the optimal tuning parameter value, over one simulated data set.} 
\label{fig:crazy}
\end{figure}

A strength of FLAM is its local adaptivity, or ability to produce a fit that is highly variable in one portion of the domain and constant in another. We can see this qualitatively by examining the function fits for scenario 4 with 96 noise functions. In Figure~\ref{fig:crazy}, we plot the fits corresponding to the the optimal tuning parameter (as defined above) for the truly non-zero functions, across 25 replicate data sets. In general, FLAM more adeptly fits both the constant and highly variable regions of the functions, relative to SpAM. SpAM's local adaptivity is limited due to the types of penalties imposed in (\ref{eq:spam}) --- $\lambda$ encourages the entire $\hat f_j$ to be zero, while $d$ controls the amount of flexibility in each $\hat f_j$. Having fewer basis functions (i.e., small $d$) results in less variable function fits, while a large $d$ can produce highly variable function fits. However, the amount of variability cannot be varied greatly over the domain of the function fit, unless basis functions are specifically chosen for this purpose \textit{a priori}.

%%%%%%%%%%%%%%%%%%%%%%%%%%%%%%%%%%%%%%%
%%%%%%%%%%%%%%%%%%%%%%%%%%%%%%%%%%%%%%%
% data application
%%%%%%%%%%%%%%%%%%%%%%%%%%%%%%%%%%%%%%%
%%%%%%%%%%%%%%%%%%%%%%%%%%%%%%%%%%%%%%%

\section{Data Application}
\label{sec:data}

\subsection{Predictors of a Country's Happiness}

We now consider  whether wealth is associated with  happiness, by estimating the conditional relationships between a country-level happiness index and gross national income, as well as 11 other country-level predictors. The happiness index is the average of Cantril Scale \citep{cantril1965pattern} responses of approximately 3000 residents in each country obtained in Gallup World Polls from 2010-2012,  publicly available from the United Nations (UN) 2013 World Happiness Report \citep{helliwell2013world}.  The predictors are publicly available through the UN Human Development Reports and the World Bank Development Indicators \citep{world2012world, undp}. They are from 2012 data or the closest year prior.

\begin{figure}[ht!]
\begin{center}
\includegraphics[width=16cm]{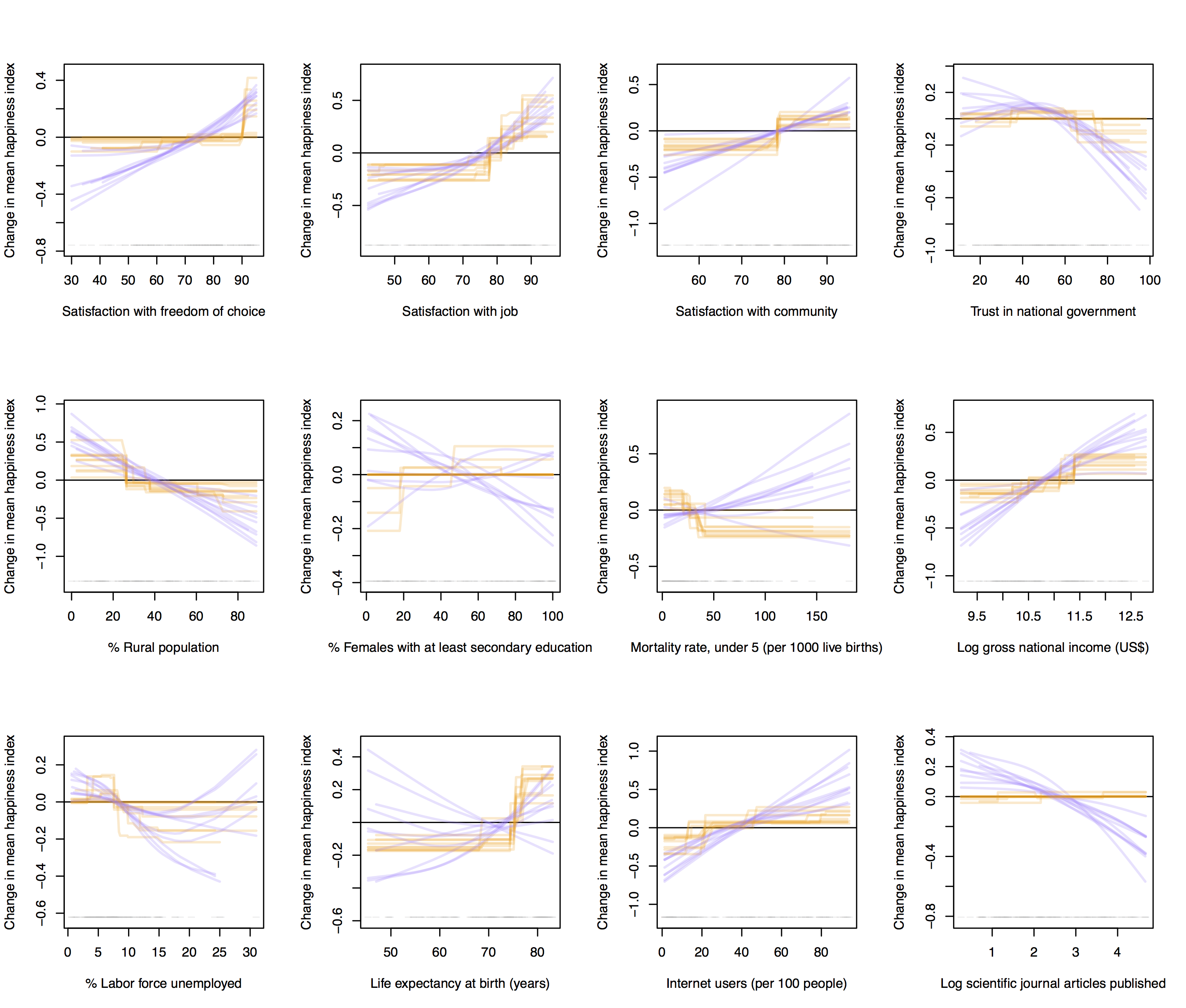}
\end{center}
\caption{Conditional associations between the happiness index for a country and twelve country-level predictors were estimated using FLAM (\protect\includegraphics[height=.15cm]{orange}) and GAM (\protect\includegraphics[height=.15cm]{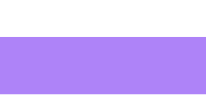}). Ten fits for each method were obtained by repeatedly splitting the data into training and test sets.  The gray bar at the bottom of each plot indicates the distribution of that predictor.}
\label{fig:happy}
\end{figure}

We consider 10 splits of the 109 countries with complete data into training and test sets. We compare FLAM to GAM with an identity link and smoothing splines using the \texttt{R} package \texttt{gam} \citep{gam}. Tuning parameters are chosen using 10-fold CV in the training set. The estimated fits are shown in Figure~\ref{fig:happy}. Both methods provide a large improvement in average test set MSE across 10 splits of the data (FLAM: 0.367; GAM: 0.308) compared to the intercept-only model (1.19). FLAM's estimated fits are both intuitive and fairly similar across the different splits of data. Conditional on the other predictors, FLAM estimates that increased gross national income is associated with increased happiness, up to a certain level of income. Beyond that, happiness is constant. We were quite surprised to see that GAM finds a \emph{negative} conditional association between a country's happiness index and the number of scientific journal publications. Reassuringly, FLAM found no such association.

\subsection{Classification based on Gene Expression}

In this section, we apply FLAM with logistic loss (to be discussed in Section~\ref{sec:ext}), in order to perform classification using gene expression measurements. The data sets we consider are:
\begin{enumerate}
\item \textit{Autism} \citep{alter2011autism}: 1498 gene expression measurements from peripheral blood lymphocytes sampled from 82 children with autism and 60 controls,  publicly available from GEO at accession number GDS4431 \citep{barrett2007ncbi}.
\item  \textit{Lung S} \citep{spira2007airway}: 22,283 gene expression measurements from large airway epithelial cells sampled from 97 smokers with lung cancer and 90 smokers without lung cancer, available from GEO at accession number GDS2771.
\item  \textit{Lung NS} \citep{lu2010identification}: 54,675 gene expression measurements from 60 pairs of tumor and adjacent normal lung tissue from non-smoking women with non-small cell lung carcinoma, available from GEO at accession number GDS3837.
\end{enumerate}
We consider only the 2000 genes with the largest variance in the \textit{Lung S} and \textit{Lung NS} data sets.  We compare the performances of FLAM to SpAM and  $\ell_1$-penalized logistic regression over 30 splits of the data into training and test sets, after standardizing each gene to have mean zero and variance one in the training set.  We choose the tuning parameters using 10-fold CV in the training set and calculate the misclassification rate in the test set.

\begin{figure}[ht!]
\begin{center}
\includegraphics[width=17cm]{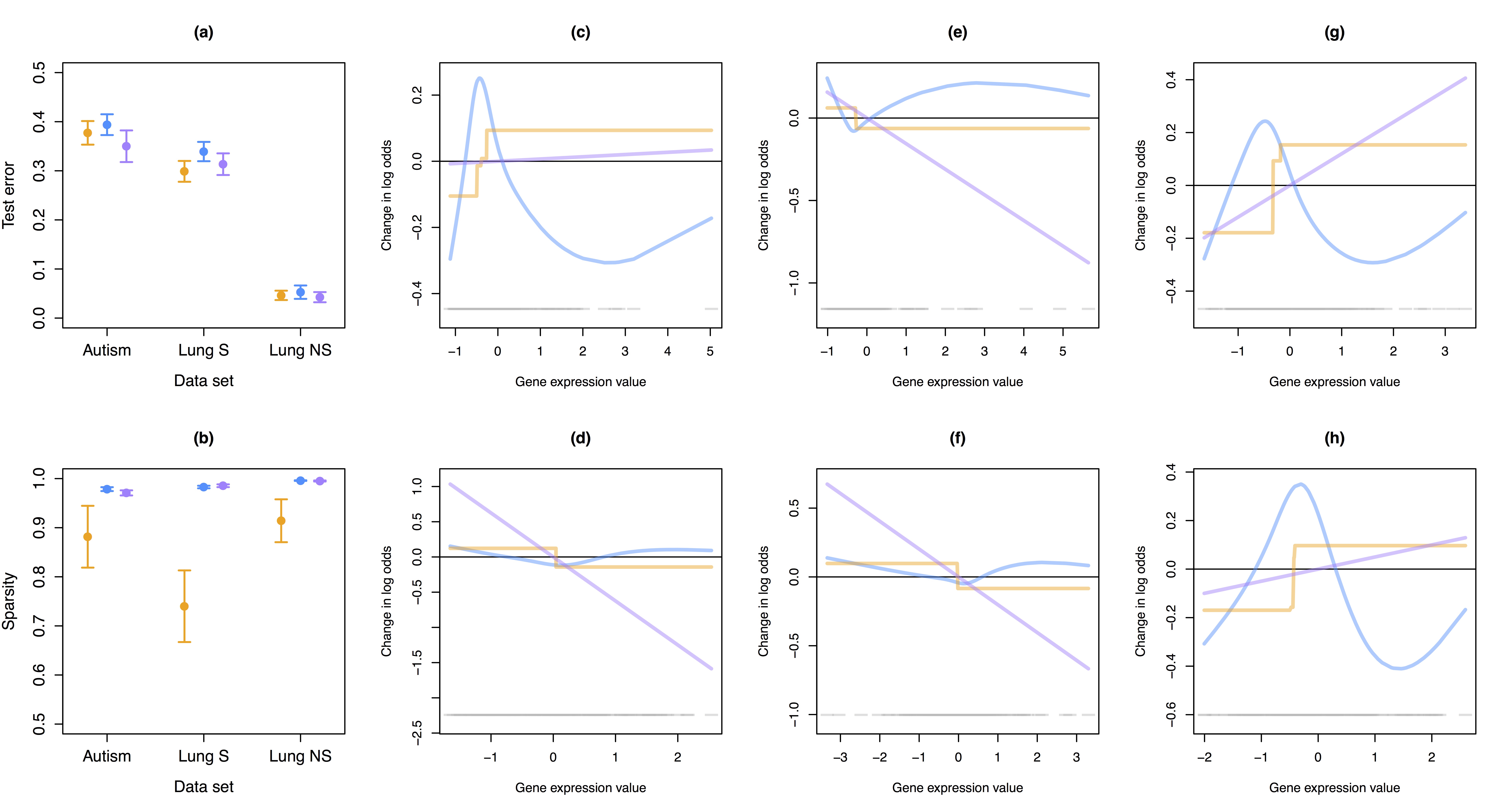}
\end{center}
\caption{We compare the classification performance of FLAM (\protect\includegraphics[height=.15cm]{orange}) to SpAM (\protect\includegraphics[height=.15cm]{blue}) and lasso (\protect\includegraphics[height=.15cm]{purple}) in terms of \textit{(a)} test set error and \textit{(b)} sparsity for three gene expression data sets. Both plots show mean estimates with 95\% confidence intervals, which are calculated using 30 splits of the data into training and test sets. In \textit{(c)}-\textit{(h)}, for six genes we show the fits from FLAM (\protect\includegraphics[height=.15cm]{orange}), SpAM (\protect\includegraphics[height=.15cm]{blue}), and lasso (\protect\includegraphics[height=.15cm]{purple}), which were estimated from one split of the Lung S data. The gray bar at the bottom of each plot indicates the distribution of that predictor.}
\label{fig:geneexp}
\end{figure}

Test error and sparsity (the percent of genes not used in the classifier) are shown in Figure~\ref{fig:geneexp}. FLAM has the same or better predictive performance on average as SpAM, but uses a less sparse classifier. However, lasso's performance is comparable to FLAM and SpAM, which indicates that the sample size may be too small to successfully model non-linear relationships in these three data sets. 

For one split of the Lung S data, Figure~\ref{fig:geneexp} displays the  estimated fits from FLAM, SpAM, and lasso for six genes. These six genes were selected because they were among the 15 with the highest-variance fits for both FLAM and SpAM.  Note that the since the genes estimated to have a non-zero relationship with the response differed for each method, the  conditional fits shown for a particular gene are not directly comparable across methods.

%%%%%%%%%%%%%%%%%%%%%%%%%%%%%%%%%%%%%%%
%%%%%%%%%%%%%%%%%%%%%%%%%%%%%%%%%%%%%%%
% extension to GAM
%%%%%%%%%%%%%%%%%%%%%%%%%%%%%%%%%%%%%%%
%%%%%%%%%%%%%%%%%%%%%%%%%%%%%%%%%%%%%%%

\section{Extensions to FLAM}
\label{sec:ext}

 We now consider the general optimization problem
\begin{equation}
\label{eq:vecform}
\displaystyle  \underset{\btheta\in\mathbb{R}^{np}}{\mathrm{minimize}} \quad \ell(\btheta) +  \lambda \sum_{j=1}^p Q_j(\btheta_j ),
\end{equation}
where $\btheta = (\btheta_1^T \; \cdots \; \btheta_p^T)^T$, $\ell:\mathbb{R}^{np}\to\mathbb{R}$ is a differentiable, convex loss function with Lipschitz continuous gradient, and $Q_j(\cdot)$ is a convex penalty function. Thus far we have considered (\ref{eq:vecform}) for squared error loss, i.e., $\ell(\btheta) = \frac{1}{2} \left \| \by-\sum_{j=1}^p \btheta_j -\theta_0\bone \right\|_2 ^2 $, and $Q_j(\btheta_j)=\alpha \left\| \bD \bP_j \btheta_j \right\|_1 + (1-\alpha) \left\| \btheta_j \right\|_2$ for $\alpha\in [0,1]$. In this section, we discuss extensions of FLAM to (1)  other losses $\ell(\btheta)$ and (2) other penalties $Q_j$. 

\subsection{A General Algorithm}
\label{subsec:genalg}

Generalized gradient descent (GGD) can be used to solve (\ref{eq:vecform}) \citep{beck2009fast}. That is, (\ref{eq:vecform}) can be solved by choosing an initial $\hat\btheta^{0}$ and continually updating
\begin{equation}
\label{eq:ggd}
\hat\btheta^{k} = \displaystyle \underset{ \btheta\in\mathbb{R}^{np}}{\mathrm{argmin}}\quad \frac{L}{2} \left\| \btheta -\hat \btheta^{k-1} + \frac{1}{L} \dot \ell (\hat\btheta^{k-1} ) \right\|_2^2+ \lambda \sum_{j=1}^p Q_j(\btheta_j )
\end{equation}
until convergence of the objective of (\ref{eq:vecform}), where $L\in\mathbb{R}$ is such that $\ddot \ell (\cdot) \preceq L\bI$. 
Equation~\ref{eq:ggd} is separable in $\btheta_j$, so the features can be updated in parallel during each iteration of GGD (in contrast to coordinate descent, in which the features are updated sequentially).
 In the special case of \eqref{eq:vecform} given in (\ref{eq:problem4}), GGD  provides an alternative to Algorithm~\ref{alg:flam}. Details are omitted in the interest of brevity. 
 
\subsection{Generalized FLAM}
\label{subsec:alggenflam}

We now consider the model $\E[y_i | \bx_i] = g\left(\sum_{j=1}^p f_j(x_{ij})\right)$, where $g(\cdot)$ is a specified function. For instance, in the case of a binary response, we can consider the mean model $\E [y_i | \bx_i] =  \text{expit}(\theta_0 + \sum_{j=1}^p \theta_{ji})$, define $\btheta = (\theta_0\bone^T\; \btheta_1^T \; \cdots \; \btheta_p^T)^T$, and take the loss to be logistic, 
\begin{equation}
\label{eq:logloss}
\ell(\btheta) = -\by^T\left( (\bone_{p+1}^T \otimes \bI_n)\btheta\right) + \bone^T  \log\left(1 + \exp\left( (\bone_{p+1}^T \otimes \bI_n)\btheta\right)\right)\nonumber.
\end{equation}
We then solve \eqref{eq:vecform} by continually updating (\ref{eq:ggd}), which amounts to the updates
$$\hat\theta_0^{k}=\hat\theta_0^{k-1} - \frac{4}{n(p+1)} \left[\text{expit}\left( \hat\theta_0^{k-1} \bone + \sum_{j=1}^p \hat\btheta^{k-1} _j\right) -  \by\right] ^T \bone$$
\small
\begin{equation}
\label{eq:partialmin2}
\hat\btheta_j^{k} =\underset{\btheta_j\in\mathbb{R}^n}{\mathrm{argmin}} \quad  \frac{p+1}{8} \left\| \btheta_j - \hat\btheta^{k-1} _j + \frac{4}{p+1} \left[\text{expit}\left( \hat\theta_0^{k-1} \bone + \sum_{j=1}^p \hat\btheta^{k-1} _j\right) -  \by\right] \right\|_2^2+ \lambda Q_j(\btheta_j ) 
\end{equation}
\normalsize for $j=1,\ldots,p$.
The solution of (\ref{eq:partialmin2}) follows from Corollary~\ref{cor:soln} when  $Q_j(\btheta_j)=\alpha \left\| \bD \bP_j \btheta_j \right\|_1 + (1-\alpha) \left\| \btheta_j \right\|_2$ for $\alpha\in [0,1]$. We now consider \eqref{eq:partialmin2} with a more general form of $Q_j$.

\subsection{FLAM with an Alternative Penalty}
\label{subsec:otherpenalties}

Thus far, we have seen that \eqref{eq:vecform} can be solved by repeatedly solving a problem of the form \eqref{eq:ggd}. When  $Q_j(\btheta_j)=\alpha \left\| \bD \bP_j \btheta_j \right\|_1 + (1-\alpha) \left\| \btheta_j \right\|_2$ for $\alpha\in [0,1]$, the solution to \eqref{eq:ggd} follows from Corollary~\ref{cor:soln}. Lemma~\ref{lem:gensoln} generalizes this Corollary to other forms of the penalty $Q_j$.
\begin{lemma}
\label{lem:gensoln}
For any norm $\| \cdot \|$, and any matrix $\bB$ with $n$ columns, the solution to 
\begin{equation}
\label{eq:geneq1}
\displaystyle \underset{\btheta\in\mathbb{R}^n}{\mathrm{minimize}} \quad  \frac{1}{2} \left \| \by -\btheta \right\|_2 ^2 + \alpha\lambda  \left\|  \bB  \btheta \right \| + (1-\alpha)\lambda \left\|   \btheta \right \|_2 
\end{equation}
is $\left(1 - \frac{(1-\alpha)\lambda}{\left\| \hat \btheta \right\|_2}\right)_+\hat\btheta$, \quad where $(u)_+= \max(u,0)$ and $\hat\btheta$ is the solution to
\begin{equation}
\label{eq:geneq2}
\displaystyle \underset{\btheta\in\mathbb{R}^n}{\mathrm{minimize}} \quad  \frac{1}{2} \left \| \by -\btheta \right\|_2 ^2 + \alpha\lambda  \left\|  \bB  \btheta \right \|.  
\end{equation}
\end{lemma}

\subsection{Simulations for Generalized FLAM using Logistic Loss} 
\label{subsec:logistic}
 
\begin{figure}[ht!]
\begin{center}
\includegraphics[width=13cm]{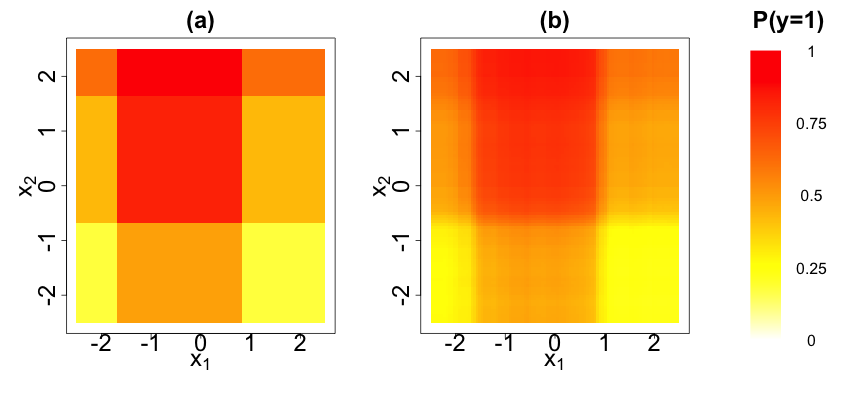}
\end{center}
\caption{We compare \textit{(a)} the true expectation of $y_i$ to \textit{(b)} the estimated expectation of $y_i$ (averaged over 25 data replicates) obtained from FLAM with logistic loss for combinations of the two predictors, $x_1$ and $x_2$.}
\label{fig:heatmap}
\end{figure}

 We now present simulation results of FLAM for logistic loss and $Q_j(\btheta_j)=\lambda \left\| \bD \bP_j \btheta_j \right\|_1$. Data are generated according to $y_i\stackrel{iid}{\sim} \text{Bernoulli}(\text{expit}[f_1 (x_{i1}) + f_2 (x_{i2}) ])$ with $x_{ij}\stackrel{iid}{\sim} \text{Uniform}[-2.5,2.5],$
where $f_1$ and $f_2$ are taken to be two of the piecewise constant functions considered previously (Figure~\ref{fig:fj}(a)). Figure~\ref{fig:heatmap}(a) shows the expectation of $y_i$ as a function of $x_1$ and $x_2$. For each replication, we generate training and test sets with $n=100$. We choose the value $\lambda$ corresponding to the minimum test set MSE. The estimated expectation of $y_i$ averaged over 25 data replicates is displayed in Figure~\ref{fig:heatmap}(b). It closely mirrors Figure~\ref{fig:heatmap}(a).

%%%%%%%%%%%%%%%%%%%%%%%%%%%%%%%%%%%%%%%
%%%%%%%%%%%%%%%%%%%%%%%%%%%%%%%%%%%%%%%
% discussion
%%%%%%%%%%%%%%%%%%%%%%%%%%%%%%%%%%%%%%%
%%%%%%%%%%%%%%%%%%%%%%%%%%%%%%%%%%%%%%%

\section{Discussion}
\label{sec:discussion}

We have presented the fused lasso additive model, a flexible yet interpretable framework for prediction, for which the estimated fits are piecewise constant with data-adaptive knots. 

While the $\ell_1$ penalty in (\ref{eq:problem4})  limits the number of knots in the fits, it also shrinks the magnitude of jumps where knots do occur. However, the resulting shrinkage can easily be addressed by debiasing the fit. That is, FLAM can be used to identify the knots; then the piecewise constant model can be refit using standard linear regression with the appropriate basis functions for the known knots. 

While the piecewise constant framework has much flexibility, a large number of knots are needed to accommodate trends with large slopes. Piecewise linear fits are more suited to this type of relationship. The problem of estimating piecewise trends of any order between a single predictor and response has been previously explored \citep{kim2009ell_1, tibshirani2014}. We leave the extension of FLAM to this setting to future work. 

The \texttt{R} package \texttt{FLAM}  will be made available on \texttt{CRAN}. The \texttt{R} package \texttt{Shiny} \citep{shiny} was used to develop interactive web applications demonstrating the performance of FLAM on simulated  and user-uploaded data (\textit{students.washington.edu/ajpete}).

%%%%%%%%%%%%%%%%%%%%%%%%%%%%%%%%%%%%%%%
%%%%%%%%%%%%%%%%%%%%%%%%%%%%%%%%%%%%%%%
% references
%%%%%%%%%%%%%%%%%%%%%%%%%%%%%%%%%%%%%%%
%%%%%%%%%%%%%%%%%%%%%%%%%%%%%%%%%%%%%%%

\footnotesize
\clearpage
\newpage
\bibliography{reference}
\bibliographystyle{plainnat}
\normalsize

%%%%%%%%%%%%%%%%%%%%%%%%%%%%%%%%%%%%%%%
%%%%%%%%%%%%%%%%%%%%%%%%%%%%%%%%%%%%%%%
% appendix
%%%%%%%%%%%%%%%%%%%%%%%%%%%%%%%%%%%%%%%
%%%%%%%%%%%%%%%%%%%%%%%%%%%%%%%%%%%%%%%

\clearpage
\newpage
\appendix

\section{Appendix}

\subsection{Proof of Lemma \ref{lem:lasso}}

Differentiating (\ref{eq:problem4}) with respect to $\theta_0$ gives $\hat\theta_0 = \frac{1}{n}\bone^T \left(\by - \sum_{j=1}^p \btheta_j\right) = \frac{1}{n}\bone^T \by$, since $\bone^T\btheta_j=0$ for $j=1,\ldots,p$. Thus we can solve \eqref{eq:problem4} with $\by$ centered and no intercept.
We now reparameterize in terms of $\bbeta_j\in\mathbb{R}^{n-1}$ where $\bbeta_j=\bD\bP_j\btheta_j$. 
Note that $$\btheta_j = \bP_j^T\bP_j \btheta_j = \bP_j^T \left(\bU\bD + \frac{1}{n}\bone\bone^T\right)\bP_j\btheta_j =\bP_j^T\bU\bbeta_j,$$ using the facts that $\bU\bD + \frac{1}{n}\bone\bone^T =\bI$ and $\bone^T\bP_j\btheta_j=0$. 
Defining $\bbeta = (\bbeta_1^T \; \cdots \; \bbeta_p^T)^T\in \mathbb{R}^{(n-1)p}$ and $\bV=[\bP_1^T \bU \cdots \bP_p^T  \bU ]$, we see that (\ref{eq:problem4}) can be rewritten as \eqref{eq:lasso}.

\subsection{Proof of Proposition~\ref{prop:df}}
We first derive the degrees of freedom for $\hat\by$ for \eqref{eq:lassoridge} when $\by\sim MVN(\bzero,\sigma^2\bI)$. Using the dual problem of \eqref{eq:lassoridge} and Lemma 1 of \citet{tibshirani2012degrees}, it can be shown that $g:\mathbb{R}^n\to\mathbb{R}^n$ with $\hat\by =g(\by)=\left(g_1(\by),\ldots,g_n(\by)\right)^T$ is continuous and almost differentiable. Thus, Stein's lemma implies that $\text{df}(\hat\by) =\E\left[ \Tr \left(\frac{\partial g(\by)}{\partial \by}\right)\right]$.
We denote the active set of $\hat\bbeta$ as $\mathcal{A}$, which is unique since (\ref{eq:lassoridge}) is strictly convex. At the optimum of (\ref{eq:lassoridge}), we have
\begin{equation}
\label{eq:zeroeqflam}
\bzero = -\bV_{\mathcal{A}}^T\left(\by - \bV_{\mathcal{A}}\hat\bbeta_{\mathcal{A}}\right) + \alpha\lambda \text{sign}\left(\hat\bbeta_{\mathcal{A}}\right) + (1-\alpha)\lambda S_1+\epsilon \hat\bbeta_{\mathcal{A}},
\end{equation}
where $S_1 = \frac{\partial}{\partial \hat\bbeta_{\mathcal{A}}} \sum_{j=1}^p\left \| \bU\bbeta_{j}\right\|_2 = \left[ S_{1,1}^T \cdots S_{1,p}^T\right]^T$ with $S_{1,j} = \frac{\bU_{\mathcal{A}_j}^T\bU_{\mathcal{A}_j}\hat\bbeta_{j\mathcal{A}_j}}{\| \bU_{\mathcal{A}_j}\hat\bbeta_{j\mathcal{A}_j}\|_2}.$

We conjecture that there is a neighborhood around almost every $\by$ (i.e., except a set of measure zero) such that $\hat\bbeta^*$ corresponding to any $\by^*$ in that neighborhood has $\mathcal{A}^*=\mathcal{A}$ and $\text{sign}(\hat\bbeta^*_{\mathcal{A}^*})=\text{sign}(\hat\bbeta_{\mathcal{A}})$. 
On the basis of this conjecture, we treat $\mathcal{A}$ and $\text{sign}(\hat\bbeta_{\mathcal{A}})$ in (\ref{eq:zeroeqflam}) as constants with respect to $\by$. Thus the derivative of (\ref{eq:zeroeqflam}) with respect to $\by$ is
\begin{equation}
\label{eq:zeroeqflam2}
\bzero = -\bV_{\mathcal{A}}^T + \bV_{\mathcal{A}}^T \bV_{\mathcal{A}}\frac{\partial\hat\bbeta_{\mathcal{A}}}{\partial\by} +(1-\alpha) \lambda S_2\frac{\partial\hat\bbeta_{\mathcal{A}}}{\partial\by} + \epsilon \frac{\partial\hat\bbeta_{\mathcal{A}}}{\partial\by},
\end{equation}
where $S_2=\frac{\partial S_1}{\partial \hat\bbeta_{\mathcal{A}}}$ is a block diagonal matrix with the $j$th block equaling
$$\frac{\bU_{\mathcal{A}_j}^T\bU_{\mathcal{A}_j}}{\| \bU_{\mathcal{A}_j}\hat\bbeta_{j\mathcal{A}_j} \|_2} - \frac{\bU_{\mathcal{A}_j}^T\bU_{\mathcal{A}_j}\hat\bbeta_{j\mathcal{A}_j}\hat\bbeta_{j\mathcal{A}_j}^T\bU_{\mathcal{A}_j}^T\bU_{\mathcal{A}_j}}{\| \bU_{\mathcal{A}_j} \hat\bbeta_{j\mathcal{A}_j} \|_2^3}.$$
Solving (\ref{eq:zeroeqflam2}) for $\frac{\partial\hat\bbeta_{\mathcal{A}}}{\partial\by}$ and left multiplying by $\bV_{\mathcal{A}}$, we have \\
$$\frac{\partial\hat\by}{\partial\by}= \bV_{\mathcal{A}}\frac{\partial\hat\bbeta_{\mathcal{A}}}{\partial\by} = \bV_{\mathcal{A}}\left[\bV_{\mathcal{A}}^T \bV_{\mathcal{A}}+ (1-\alpha)\lambda S_2+ \epsilon\bI\right]^{-1} \bV_{\mathcal{A}}^T.$$
Therefore, the degrees of freedom are $\E \left[\Tr\left( \bV_{\mathcal{A}}\left[ \bV_{\mathcal{A}}^T \bV_{\mathcal{A}}+ (1-\alpha)\lambda S_2+ \epsilon\bI\right]^{-1}  \bV_{\mathcal{A}}^T  \right)\right].$
This yields the estimator (\ref{eq:dfestflam}), where one degree of freedom is added for  the intercept.

 The proof of Proposition~\ref{prop:dfspam} is omitted, as it follows the arguments in this proof closely.

\subsection{Proof of Lemma \ref{lem:sparse}}

 The optimality condition for (\ref{eq:lasso}) with $\alpha=1$ is $-\bV^T(\tilde\by-\bV\bbeta) + \lambda s(\bbeta) =0$, where $s(\bbeta_j)=\text{sign}(\bbeta_j)$ if $\bbeta_j\neq 0$ and $s(\bbeta_j)\in[-1,1]$ if $\bbeta_j=0$. After plugging in $\bbeta=\bzero$, we obtain $-\bV^T\tilde\by + \lambda s(\bzero)=\bzero$, which is satisfied if and only if $\lambda\geq\| \bV^T\tilde\by\|_\infty$ since $s(\bzero)\in[-1,1]$.

Now we consider the optimality condition for (\ref{eq:problem4}) when $\alpha=0$, which takes the form $-\tilde\by+\sum_{j=1}^p \btheta_j  + \lambda s(\btheta_j)=\bzero$ for $j=1,\ldots,p ,$ where $s(\btheta_j)=\btheta_j/\| \btheta_j \|_2$ if $\btheta_j \neq \bzero$ and $s(\btheta_j)\in\{ \bg  \mid \| \bg\|_2\leq 1\}$ if $\btheta_j=\bzero$. After plugging in $\btheta_j=\bzero$ for $j=1,\ldots,p$, we obtain $-\tilde\by  + \lambda s(\bzero)=\bzero$, which is satisfied if and only if $\lambda\geq \| \tilde\by\|_2$.

\subsection{Proof of Corollary \ref{cor:sparseall}}

The  objective of (\ref{eq:problem4}) is bounded below by 
\begin{equation}
\label{eq:bound1}
 \underset{\theta_0\in\mathbb{R}, \btheta_j\in\mathbb{R}^n, 1\leq j \leq p}{\mathrm{min}} \quad  \frac{1}{2} \left \| \by-\sum_{j=1}^p \btheta_j - \theta_0 \bf{1} \right\|_2 ^2 + \alpha\lambda \sum_{j=1}^p \left\|  \bD \bP_j \btheta_j \right \|_1 
\end{equation}
and
\begin{equation}
\label{eq:bound2}
\underset{\theta_0\in\mathbb{R}, \btheta_j\in\mathbb{R}^n, 1\leq j \leq p}{\mathrm{min}} \quad  \frac{1}{2} \left \| \by-\sum_{j=1}^p \btheta_j - \theta_0 \bf{1} \right\|_2 ^2 + (1-\alpha)\lambda \sum_{j=1}^p \left\|   \btheta_j \right \|_2
\end{equation}
for any $\alpha\in [0,1]$ and $\lambda\geq 0$. 
Lemma~\ref{lem:sparse} implies that if 
 $\lambda \geq \min \left( \frac{\|  \bV^T\tilde\by \|_\infty}{\alpha}, \frac{\| \tilde\by\|_2}{1-\alpha} \right),$ then the  objective of (\ref{eq:problem4}) is bounded below by $\underset{\theta_0\in\mathbb{R}}{\mathrm{min}} \quad \frac{1}{2} \left \| \by- \theta_0 \bf{1} \right\|_2 ^2 = \frac{1}{2} \left \| \by- \frac{1}{n}\bone^T\by \bf{1} \right\|_2 ^2$. The objective of (\ref{eq:problem4}) achieves this lower bound when $\hat\btheta_j=\bzero$ for $j=1,\ldots,p$.

\subsection{Proof of Lemma \ref{lem:consist}}

\begin{fact}
\label{fact:normal}
Let $\bz_j\in\mathbb{R}^n\sim MVN(\bzero,\Sigma)$ with $\underset{1\leq i\leq n}{\max} \Sigma_{i,i} \leq c$ for $j=1,\ldots,p$. Then $$P\left(\underset{1\leq j \leq p}{\max}\| \bz_j \|_\infty \geq 2\sqrt{c\log (np)}\right)\leq \frac{2}{np}.$$
\end{fact}
\textit{Proof:}
Note that $P(z_{ji}\geq 2\sqrt{c\log (np)}) \leq P(z_0 \geq 2\sqrt{c\log (np)})$ where $z_0\sim N(0,c)$. Thus
$$P\left(\underset{1\leq j \leq p}{\max}\| \bz_j\|_\infty \geq 2\sqrt{c\log (np)}\right)\leq 2npP\left( z_0/\sqrt{c} \geq 2\sqrt{\log (np)}\right)\leq\frac{2}{np},$$
which follows from the union bound and the fact that 
$$P\left( z_0/\sqrt{c} \geq 2\sqrt{\log (np)}\right)  = \int_{2\sqrt{\log (np)}}^\infty \frac{1}{\sqrt{2\pi}} \exp\left(\frac{-t^2}{2}\right) dt \leq   \int_{2\sqrt{\log (np)}}^\infty t \exp\left(\frac{-t^2}{2}\right) dt= \frac{1}{n^2p^2}.$$

\begin{fact}
\label{fact:chi}
Let $\bz\in\mathbb{R}^n\sim MVN(\bzero,c\bI)$. Then \small$P\left(\| \bz \|_2 \geq \sqrt{cn\left( 1+\sqrt{\frac{4\log n}{n}} + \frac{4\log n}{n}\right)} \right)\leq \frac{1}{n}.$\normalsize
\end{fact}
\textit{Proof:}
Since $\| \bz \|_2^2 \sim c\chi_n^2$, this follows from Lemma 8.1 in \citet{buhlmann2011statistics}.

We now establish prediction consistency. We rewrite \eqref{eq:problem4} as $$\underset{\btheta\in\mathbb{R}^{np}}{\mathrm{minimize}} \quad  \frac{1}{2n} \left \| \by-\bW\btheta \right\|_2 ^2 + \lambda \Omega(\btheta),$$
where $\bW=\bone^T\otimes \bI_n$ and $\Omega(\btheta) = \sum_{j=1}^p \left[ \alpha\left\|  \bD \bP_j \btheta_j \right \|_1 + (1-\alpha)\|\btheta_j\|_2 \right]$. Denote the true coefficient vector as $\btheta^0 = (\btheta_1^{0T} \cdots \btheta_p^{0T})^T$ and assume $\by = \bW\btheta^0 + \bepsilon$ with $\bepsilon\sim MVN(\bzero,\sigma^2\bI)$. By the definition of $\hat\btheta$, $\frac{1}{2n} \left \| \by-\bW\hat\btheta \right\|_2 ^2 + \lambda \Omega(\hat\btheta) \leq \frac{1}{2n} \left \| \by-\bW\btheta^0 \right\|_2 ^2 + \lambda \Omega(\btheta^0),$ so 
\begin{equation}
\label{eq:main}
\frac{1}{2n} \left \| \sum_{j=1}^p (\hat\btheta_j-\btheta_j^0) \right\|_2 ^2 + \lambda \Omega(\hat\btheta) \leq V_n(\hat\btheta) + \lambda \Omega(\btheta^0)
\end{equation}
where $V_n(\hat\btheta)=\frac{1}{n}\sum_{j=1}^p \bepsilon^T(\hat\btheta_j-\btheta_j^0).$ We wish to bound the empirical process $V_n(\hat\btheta)$. We have
\begin{align}
\left\vert V_n(\hat\btheta)\right\vert &= \frac{1}{\sqrt{n}} \left\vert\sum_{j=1}^p \left[ \frac{\alpha}{\sqrt{n}}\bepsilon^T\bP_j^T\bU\bD\bP_j(\hat\btheta_j - \btheta_j^0) +\frac{1-\alpha}{\sqrt{n}}\bepsilon^T(\hat\btheta_j - \btheta_j^0)  \right] \right\vert\nonumber\\
&\leq \frac{1}{\sqrt{n}} \sum_{j=1}^p \left[ \left\vert\frac{\alpha}{\sqrt{n}}\bepsilon^T\bP_j^T\bU\bD\bP_j(\hat\btheta_j - \btheta_j^0) \right\vert+\left\vert\frac{1-\alpha}{\sqrt{n}}\bepsilon^T(\hat\btheta_j - \btheta_j^0)  \right\vert\right] \nonumber\\
&\leq \frac{1}{\sqrt{n}} \sum_{j=1}^p \left[ \| \bv_{1j}\|_\infty \alpha\|\bD\bP_j(\hat\btheta_j - \btheta_j^0)\|_1 +\|\bv_2\|_2 (1-\alpha)\|\hat\btheta_j - \btheta_j^0\|_2  \right] \nonumber\\
&\leq \frac{ \underset{1\leq j \leq p}{\max}  \| \bv_{1j}\|_\infty}{\sqrt{n}}\sum_{j=1}^p \alpha\left(\|\bD\bP_j\hat\btheta_j\|_1 +  \|\bD\bP_j\btheta_j^0\|_1\right)  +\frac{\|\bv_2\|_2}{\sqrt{n}} \sum_{j=1}^p (1-\alpha)\left(\|\hat\btheta_j\|_2 +  \|\btheta_j^0\|_2\right) \nonumber
\end{align}
where $\bv_{1j} = \frac{1}{\sqrt{n}} \bU^T\bP_j\bepsilon$ and $\bv_2=\frac{1}{\sqrt{n}}\bepsilon$. 

We now establish bounds for $\underset{1\leq j \leq p}{\max}\|\bv_{1j}\|_\infty$ and $\|\bv_2\|_2$ that hold with large probability.

 Note that $\bv_{1j}\in\mathbb{R}^{n-1}\sim MVN(\bzero,\frac{\sigma^2}{n}\bU^T\bU)$. Since
$\underset{1\leq i \leq n}{\max} ( \frac{\sigma^2}{n}\bU^T\bU)_{i,i} \leq \frac{\sigma^2}{4}$,
Fact~\ref{fact:normal} with $c=\frac{\sigma^2}{4}$ gives
$P\left(\underset{1\leq j \leq p}{\max}\frac{\| \bv_{1j} \|_\infty}{\sqrt{n}} \geq w_1\right)\leq \frac{2}{(n-1)p}$
where $w_1=\sigma\sqrt{\frac{\log ((n-1)p)}{n}}$.  
Now recall $\bv_2\sim MVN(\bzero,\frac{\sigma^2}{n}\bI)$, so by Fact~\ref{fact:chi},
$P\left(\frac{\| \bv_2 \|_2}{\sqrt{n}} \geq w_2\right)\leq \frac{1}{n}$
where $w_2=\frac{\sigma}{\sqrt{n}}\sqrt{1 + \sqrt{\frac{4\log n}{n}} + \frac{4\log n}{n}}$.

Therefore, with probability at least $1-\left(\frac{2}{(n-1)p} + \frac{1}{n}\right)$,
$$\vert V_n(\hat\btheta)\vert\leq \sum_{j=1}^p [ w_1 \alpha(\|\bD\bP_j\hat\btheta_j\|_1 +  \|\bD\bP_j\btheta_j^0\|_1)  + w_2 (1-\alpha)(\|\hat\btheta_j\|_2 +  \|\btheta_j^0\|_2)].$$
Let $\lambda= 2 w_1$. For $p\geq 1$ and $n\geq 15$, $w_1>w_2$. Thus, with high probability, \small
\begin{equation}
\label{eq:bound}
\vert V_n(\hat\btheta)\vert \leq\frac{\lambda}{2} \sum_{j=1}^p[ \alpha (\|\bD\bP_j\hat\btheta_j\|_1 +  \|\bD\bP_j\btheta_j^0\|_1) + (1-\alpha) (\|\hat\btheta_j\|_2+ \|\btheta_j^0\|_2)  ]= \frac{\lambda}{2} (\Omega(\hat\btheta) + \Omega(\btheta^0)). 
\end{equation}
\normalsize
The result follows from plugging the bound from \eqref{eq:bound} into \eqref{eq:main}.

\subsection{Derivations of Results from Section~\ref{subsec:alggenflam}}
\label{subsec:pfcor2}
For $\ell(\btheta)$ of the form \eqref{eq:logloss}, we can calculate
\begin{eqnarray*}
\dot \ell (\btheta)  &=& (\bone_{p+1} \otimes \bI_n) \left[\text{expit}( (\bone_{p+1}^T \otimes \bI_n)\btheta) -  \by\right] \\
\ddot \ell (\btheta) &=&(\bone_{p+1} \otimes \bI_n) \text{diag}\left(\text{expit}( (\bone_{p+1}^T \otimes \bI_n)\btheta) (1-\text{expit}( (\bone_{p+1}^T \otimes \bI_n)\btheta) )\right) (\bone_{p+1}^T \otimes \bI_n)\\ 
 &\preceq&  \frac{1}{4}(\bone_{p+1}\bone_{p+1}^T \otimes \bI_n) \preceq  \frac{1}{4}(p+1)\bI.
\end{eqnarray*}
Thus, plugging in  $L= \frac{1}{4}(p+1)$ and $\dot\ell(\btheta)$ into (\ref{eq:ggd}), we now have 
\small
\begin{equation}
\label{eq:logistic}
\hat\btheta^{k} = \displaystyle \underset{ \btheta\in\mathbb{R}^{n(p+1)}}{\mathrm{argmin}}\quad \displaystyle  \frac{p+1}{8} \left\| \btheta - \hat\btheta^{k-1}  + \frac{4}{p+1} (\bone_{p+1} \otimes \bI_n) \left[\text{expit}( (\bone_{p+1}^T \otimes \bI_n)\hat\btheta^{k-1} ) -  \by\right] \right\|_2^2+ \lambda \sum_{j=1}^p Q_j(\btheta_j ). \nonumber
\end{equation}
\normalsize
This is separable in  $\btheta_j$, and is equivalent to \eqref{eq:partialmin2} by inspection.

\subsection{Proof of Lemma \ref{lem:gensoln}}
There are two main tasks: 
\begin{list}{}{}
\item{Task 1:} Derive the form of $\hat\btheta$, the solution to (\ref{eq:geneq2}).
\item{Task 2:} Show that the solution to (\ref{eq:geneq1}) is $\tilde\btheta=(1 - (1-\alpha)\lambda/\| \hat \btheta \|_2)_+\hat\btheta$. 
\end{list}
We begin with Task 1. We rewrite \eqref{eq:geneq2} as
$$\underset{\btheta,\bz}{\mathrm{minimize}} \quad\frac{1}{2} \left \| \by -\btheta \right\|_2 ^2 + \alpha\lambda  \left\|  \bz \right \| \text{ subject to }\bz=\bB\btheta,$$
which has Lagrangian $\mathcal{L}(\btheta,\bz,\bv) = \frac{1}{2} \left \| \by -\btheta \right\|_2 ^2 + \alpha\lambda  \left\|  \bz \right \| + \bv^T(\bB\btheta - \bz).$
The dual function is $$g(\bv)=\underset{\btheta,\bz}{\text{inf}}\enskip  \mathcal{L}(\btheta,\bz,\bv) = \underset{\bz}{\text{inf}}    \left\{ \frac{1}{2} \left \| \bB^T\bv \right\|_2 ^2 + \bv^T  {\bB} (\by-\bB^T\bv) + \alpha\lambda  \left\|  \bz \right \|  -\bv^T \bz \right\} ,$$ where the second equality follows from noting that the partial minimum with respect to $\btheta$ satisfies $\btheta=\by-\bB^T\bv$.   
 Thus $g(\bv) = \frac{1}{2} \left \| \bB^T\bv \right\|_2 ^2 + \bv^T {\bB} (\by-\bB^T\bv)$ if $\| \bv\|_*\leq \alpha\lambda$ and $-\infty$ otherwise,
where $\| \cdot\|_*$ is the dual norm of $\|\cdot\|$. Finally, the dual problem is 
$$\underset{ \| \bv \|_* \leq \alpha\lambda}{\mathrm{maximize}}\quad  -\frac{1}{2} \left \| \by -\bB^T\bv  \right\|_2 ^2 +\frac{1}{2}\by^T\by.$$
Letting $\hat\bv = \underset{ \| \bv \|_* \leq \alpha\lambda}{\mathrm{argmin}}\quad \left \| \by -\bB^T\bv  \right\|_2 ^2$, the solution to (\ref{eq:geneq2}) is $\hat\btheta = \by -\bB^T\hat\bv$.

We now move on to Task 2.
Rewriting \eqref{eq:geneq1} as 
$$ \underset{\btheta,\bz_1,\bz_2}{\mathrm{minimize}} \quad  \frac{1}{2} \left \| \by -\btheta \right\|_2 ^2 + \alpha\lambda  \left\|  \bz_1 \right \|+(1-\alpha)\lambda  \left\|  \bz_2 \right \|_2 \text{ subject to }\bz_1=\bB\btheta, \bz_2=\btheta$$
and writing out the Lagrangian, one can show that 
the dual problem is 
\begin{equation}
\underset{ \| \bv \|_* \leq \alpha\lambda, \| \bu \|_2 \leq (1-\alpha)\lambda}{\mathrm{maximize}}\quad  \frac{1}{2} \left \| \bB^T\bv+\bu \right\|_2 ^2 + (\bB^T\bv+\bu)^T (\by - \bB^T\bv - \bu);
\label{dualeq}
\end{equation}
the calculations to obtain \eqref{dualeq} indicate that $\btheta=\by-\bB^T\bv-\bu$.
Problem \eqref{dualeq} is equivalent to  $\underset{ \| \bv \|_* \leq \alpha\lambda, \| \bu \|_2 \leq (1-\alpha)\lambda}{\mathrm{minimize}}\quad  \frac{1}{2} \left \| \by - \bB^T\bv - \bu \right\|_2 ^2.$
Minimizing in $\bu$, we have 
\begin{center}
$\tilde\bu = \begin{cases}
\by - \bB^T\bv & \text{if }\left \| \by - \bB^T\bv \right\|_2 \leq(1-\alpha)\lambda\\
(1-\alpha)\lambda\frac{ \by - \bB^T\bv}{\left \| \by - \bB^T\bv \right\|_2} & \text{if }\left \| \by - \bB^T\bv \right\|_2 >(1-\alpha)\lambda
\end{cases}$,
\end{center}
 the projection of $\by - \bB^T\bv $ onto the $(1-\alpha)\lambda$ ball. Thus \eqref{dualeq} is equivalent to  $$\underset{ \| \bv \|_* \leq \alpha\lambda}{\mathrm{minimize}}\quad   \left( \left \| \by - \bB^T\bv \right\|_2 - (1-\alpha)\lambda\right)_+,$$
which is solved by $\tilde\bv = \underset{ \| \bv \|_* \leq \alpha\lambda}{\mathrm{argmin}}\quad \left \| \by -\bB^T\bv  \right\|_2 ^2$. Therefore, we have shown that $\tilde\bv=\hat\bv$ and $\tilde\btheta=\by - \bB^T\tilde\bv - \tilde\bu$. It follows that $\tilde\btheta = \hat\btheta- \tilde\bu =\left(1-\frac{(1-\alpha)\lambda}{\|\hat\btheta\|_2} \right)_+ \hat\btheta.$
\end{document}